\documentclass[a4paper,11pt]{article}
\pdfoutput=1 

\usepackage{jcappub} 

 \pdfoutput=1
\usepackage[utf8]{inputenc}
\usepackage{epsfig}
\usepackage{amsmath}
\usepackage{slashed} 
\usepackage{amsfonts} 
\usepackage{float} 
\usepackage{amssymb}
\usepackage{color}
\usepackage{pbox}
\usepackage{subfigure}
\usepackage{array,multirow,makecell} 
\usepackage{tabularx}
\usepackage{titlesec} 
\usepackage{natbib}
\usepackage{lipsum}

\def\lsim{\raise0.3ex\hbox{$\;<$\kern-0.75em\raise-1.1ex\hbox{$\sim\;$}}}
\def\gsim{\raise0.3ex\hbox{$\;>$\kern-0.75em\raise-1.1ex\hbox{$\sim\;$}}}
\def\be{\begin{equation}}
\def\ee{\end{equation}}
\def\bea{\begin{eqnarray}} 
\def\eea{\end{eqnarray}}
\def\nn{\nonumber}
\DeclareUnicodeCharacter{202F}{\,}

\newcommand{\abs}[1]{\left| #1 \right|}
\newcommand{\beq}{\begin{equation}}
\newcommand{\eeq}{\end{equation}}
\newcommand{\beqn}{\begin{eqnarray}}
\newcommand{\eeqn}{\end{eqnarray}}

\def\lsim{\raise0.3ex\hbox{$\;<$\kern-0.75em\raise-1.1ex\hbox{$\sim\;$}}}
\def\gsim{\raise0.3ex\hbox{$\;>$\kern-0.75em\raise-1.1ex\hbox{$\sim\;$}}}
\def\be{\begin{equation}}
\def\ee{\end{equation}}
\def\bea{\begin{eqnarray}}
\def\eea{\end{eqnarray}}
\def\nn{\nonumber}

\titleformat{\paragraph}
{\normalfont\normalsize\bfseries}{\theparagraph}{1em}{}
\titlespacing*{\paragraph}
{0pt}{3.25ex plus 1ex minus .2ex}{1.5ex plus .2ex}

\begin{document}

\begin{flushright}
HRI-RECAPP-2021-012
\end{flushright} 
 
\title{\boldmath Self-interacting freeze-in dark matter in a singlet doublet scenario}


\author[a]{Purusottam Ghosh,}
\author[b]{Partha Konar,}
\author[c]{Abhijit Kumar Saha,} 
\author[b,d]{and Sudipta Show}


\affiliation[a]{Regional Centre for Accelerator-based Particle Physics,
Harish-Chandra Research Institute, HBNI,
Chhatnag Road, Jhunsi, Allahabad - 211 019, India}
\affiliation[b]{Theoretical Physics Division, Physical Research Laboratory, Shree Pannalal Patel Marg, Ahmedabad - 380009, Gujarat, India}
\affiliation[c]{School of Physical Sciences, Indian Association for the Cultivation of Science, 2A $\&$ 2B Raja S.C. Mullick Road, Kolkata 700 032, India}
\affiliation[d]{Indian Institute of Technology, Gandhinagar - 382424, Gujarat, India}

\emailAdd{purusottamghosh@hri.res.in}
\emailAdd{konar@prl.res.in}
\emailAdd{psaks2484@iacs.res.in}
\emailAdd{sudipta@prl.res.in}

\abstract{We examine the non-thermal production of dark matter in a scalar extended singlet doublet fermion model where the lightest admixture of the fermions constitutes a suitable dark matter candidate. The dark sector is non-minimal with the MeV scale singlet scalar, which is stable in the Universe lifetime and can mediate the self-interaction for the multi-GeV fermion dark matter mitigating the small scale structure anomalies of the Universe. If the dark sector is strongly coupled, it undergoes internal dark thermal equilibrium after freeze-in production, and we end up with suppressed relic abundance for the fermion dark matter in a radiation dominated Universe. In contrast, the presence of a modified cosmological phase in the
early era drives the fermion dark matter to satisfy nearly the whole amount
of observed relic.
 It also turns out that the assumption of an unconventional cosmological history can allow the GeV scale dark matter to be probed at LHC from displaced vertex signature with improved sensitivity.}


\maketitle

\flushbottom

\section{Introduction}
In the sea of apparent voidness at experimental fronts, some remarkable signs of progress have recently been made in dark matter theory, with new elegant ideas concerning dark matter production mechanisms and the nature of its interaction. One of the exciting proposals is the non-thermal origin of dark matter production known as an {\it freeze-in} \cite{Hall:2009bx}. The freeze-in process holds under the assumption 
that the dark matter is feebly coupled to the visible sector and initially having zero number density. At a later stage, it can be produced non thermally from decay or scattering of visible sector particles.

On another front, the collisionless cold dark matter paradigm suffers from small scale structure anomalies such as {\it cusp vs core} \cite{1991ApJ...378..496D, Flores:1994gz, Tulin:2013teo} and the {\it too-big-to-fail} problems \cite{2011, 2010, Tulin:2013teo} \textit{etc}. As a resolution it has been proposed that the self-interaction \cite{Spergel:1999mh,Bullock:2017xww, Klypin:1999uc, Moore:1999nt, Moore:1999gc, Springel:2008cc}, a form of non-gravitational interactions of Dark Matter (DM), is capable of addressing the problems associated with the small scale structure. Tight constraints on dark matter self-interactions come in terms of
the ratio of such self-scattering cross section over DM mass from
observations at Bullet
Cluster and N-body simulations \cite{Randall:2008ppe, Tulin:2013teo}. 

 Studies related to self-interacting dark matter in a typical WIMP (weakly interacting massive particle) paradigm 
are prevalent in literature (see, for example, \cite{Kouvaris:2014uoa, Kainulainen:2015sva,Borah:2021jzu,Dutta:2021wbn,Borah:2021yek} for some of the recent works). On the other hand, proposals for self-interacting dark matter with freeze-in production are somewhat limited and can be found in refs. \cite{Duch:2017khv, Du:2020avz}. In ref. \cite{Du:2020avz}, the dark matter is fermionic in nature, and it self-interacts via a light scalar mediator. Similar analysis has been exercised in ref. \cite{Duch:2017khv}, considering a vector dark matter candidate. Irrespective of the production mechanisms, a strong dark matter self-interaction craves for a light mediator with mass ranging from keV to MeV scale. Another notable aspect is the dark sector thermalisation \cite{Du:2020avz} due to the strong interaction between the dark matter and the light mediator. This reduces the relic abundance of DM after its freeze-in production. To avoid that, restriction on the mediator mass turns severe and lowers it towards $\mathcal{O}(1)$ KeV.

In this paper, we attempt to offer a freeze-in production \cite{Hall:2009bx} scenario of GeV scale vector fermion DM that carries a bright prospect of testability at collider experiments and is capable of alleviating small scale structure problems by undergoing sufficient self-interaction. In view of this, we adopt the singlet doublet framework where the dark matter ($\chi$) turns out to be the lightest admixture of the singlet and doublet \cite{DEramo:2007anh, Enberg:2007rp, Cohen:2011ec, Cheung:2013dua, Restrepo:2015ura, Calibbi:2015nha, Bhattacharya:2015qpa, Yaguna:2015mva, Horiuchi:2016tqw, Banerjee:2016hsk, Abe:2017glm, Maru:2017pwl, Maru:2017otg, Bhattacharya:2017sml, Xiang:2017yfs, Esch:2018ccs, Arcadi:2018pfo, DuttaBanik:2018emv, Bhattacharya:2018fus, Fiaschi:2018rky, Barman:2019tuo, Restrepo:2019soi, Barman:2019aku, Konar:2020wvl, Konar:2020vuu, Dutta:2020xwn, Barman:2021ifu, Borah:2021khc} \footnote{Earlier works \cite{Calibbi:2018fqf,No:2019gvl} on singlet doublet freeze-in dark matter reveal that the dark matter is mostly singlet dominated with vanishing mixing angle with the doublet.}. We extend the set-up by adding one gauge singlet real scalar ($\phi$) field. The new scalar mediates the self-interaction for the fermion dark matter candidate. Additionally, it has no decay mode and is hence stable in Universe lifetime. The singlet doublet dark matter model has rich collider signatures as explored in \cite{Calibbi:2015nha, Calibbi:2018fqf, No:2019gvl}. The heavier charged fermion or neutral partner is part of the additional $SU(2)_L$ doublet fermion.  Their copious production at colliders is expected due to the $SU(2)_L$ gauge mediated interactions. The long-lived charged particle could give rise to disappearing charge tracks, whereas the long-lived heavier neutral particle can be traced through searching displaced vertices at LHC.

We focus on working in a scenario where the GeV scale fermion dark matter predominantly contribute to the total relic abundance. Stronger interaction strength between the fermionic dark matter and the mediator scalar is desired for producing large self-interaction \cite{Kouvaris:2014uoa, Kainulainen:2015sva, Zhu:2021pad, Duch:2017khv, Du:2020avz}. {On the other hand, a strongly coupled dark sector generally leads to the formation of internal dark thermal equilibrium with a separate dark sector temperature $T_D$ \cite{Du:2020avz, Krnjaic:2017tio, Berger:2018xyd, Cheung:2010gj, Cheung:2010gk, Chu:2011be}. In our present scenario, after the freeze-in production of $\chi$, they start annihilating into a pair of $\phi$ and vice versa, maintaining the dark sector equilibrium before a freeze-out mechanism triggers for $\chi$.
In literature, such rich dynamics inside a hidden dark sector is familiarly known as the reannihilation effect. Considering radiation dominated Universe (RD), we find that after the dark sector freezes out, $\phi$ shares a sizeable contribution to the total relic abundance. Thus, a $\chi$ dominated scenario is very unlikely.}

Since our aim is to provide a $\chi$ dominated scenario which may also give rise to
velocity-dependent fermionic self-interacting dark matter (see \cite{Du:2020avz} for a similar exercise in a different framework), the proposed set up as discussed above in the radiation dominated Universe does not work. In the present study, we apprehend that the presence of a non-standard epoch in the early Universe could be useful to alleviate this. A faster expansion of the Universe can prevent the dark sector from reaching thermal equilibrium or slow down the dark sector interaction rate. These assist in suppressing the $\chi\chi\to\phi\phi$ conversion rate to obtain a $\chi$ dominated dark matter scenario. We have employed two particular kinds of modified cosmology \cite{DEramo:2017gpl, Konar:2020vuu, Barman:2021ifu} namely kination domination and faster than kination domination. We have found that such non-standard epochs in the early Universe can work remarkably to reach our defined aim of obtaining $\Omega_\chi h^2\sim 0.12$ \cite{Planck:2018vyg} which has the potential to resolve the cosmological problems as earlier stated. Additionally, the presence of a non-standard epoch helps to provide a complementary probe of the GeV scale dark matter in the singlet doublet framework via displaced vertices which is not possible in a radiation dominated scenario as shown in \cite{Calibbi:2018fqf, No:2019gvl}.

The paper is organised as follows. In section \ref{model}, we furnish the model structure and the corresponding Lagrangians. Next, in section \ref{DMPH}, we discuss the dark matter phenomenology considering the standard cosmology and two different forms of non-standard cosmology. We describe the  discovery prospects of the proposed model at collider experiments in section \ref{CC}. Estimate of the self-interaction cross section for the singlet doublet dark matter are presented in section \ref{SI}.Finally, we summarise by pointing out the new findings of our analysis and draw the conclusion in section \ref{conclusion}.

\section{The Model}\label{model}
We extend the SM particle content by one $SU(2)_L$ doublet fermion ($\Psi$) with hypercharge $Y=-\frac{1}{2}$ and one hyperchargeless gauge singlet vector fermion ($\chi$). In addition, we also include a real scalar singlet field ($\phi$) which assists the DM to yield strong dark matter self-interaction. We impose a discrete $Z_2$ symmetry under which the SM particles and the $\phi$ field transform trivially while the BSM fermions are assigned odd $Z_2$ charges. The Lagrangian of the scalar sector is read as
\begin{align}
\mathcal{L}_{\rm scalar}=|D^\mu H|^2+\frac{1}{2}(\partial^\mu\phi)(\partial_\mu\phi)-V(H,\phi),
\end{align}
where the covariant derivative is defined as,
\begin{align}
D^\mu= \partial^\mu-ig\frac{\sigma^a}{2}W^{a\mu}-ig^\prime\frac{Y}{2}B^\mu,
\end{align}
with $g$ and $g^\prime$ being the $SU(2)_L$ and the $U(1)_Y$ gauge couplings respectively.
The scalar potential $V=V(H)+V(\phi)+V(\phi,H)$ takes the following form,
\begin{align}
& V(H)=-{\mu_H^2}\, (H^\dagger H)+\lambda_H \, (H^\dagger H)^2 ,\\
& V(\phi)= \frac{m_\phi^2}{2}\phi^2+\frac{\lambda_\phi}{4!}\phi^4+\frac{b_3}{3!}\phi^3,\\
& V(\phi,H)=\frac{\lambda_{\phi H}}{2}\phi^2(H^\dagger H)+a_3~\phi(H^\dagger H).
\end{align}
Note that we do not write the liner term for $\phi$ in $V(\phi)$ since it can always be absorbed through the redefinition of the other parameters. We assume that all the mass scales and the coupling coefficients are real and positive. In this limit, the vacuum expectation values (vev) of the scalars $H$ and $\phi$ after minimising the potential $V$ are obtained as,
\begin{align}\label{eq:vac}
\langle H\rangle=v ~, ~~~\langle\phi\rangle= 0.
\end{align} 
The Lagrangian for the fermionic sector is written as,
\begin{align}
\mathcal{L}=\mathcal{L}_{f}+\mathcal{L}_{Y},
\end{align}
where,
\begin{align}
\mathcal{L}_{f}=& \; i\overline{\Psi} \gamma_\mu D^\mu \Psi+i\overline{\chi} \gamma_\mu\partial^\mu \chi
-m_\Psi\overline{\Psi} \Psi
-m_\chi \overline{\chi} \chi\label{eqn:Fyukawa}\\
\mathcal{L}_{Y}=& ~-Y\overline{\Psi} \tilde{H} \chi+h.c.-\lambda\phi\overline{\chi}\chi -\delta \phi\overline{\Psi}\Psi ,\label{eqn:Fyukawa}
\end{align}
where we have defined $\Psi^T=(\psi^+ ~~\psi^0)$.

The Dirac mass matrix for the neutral fermions after the spontaneous breakdown of the electroweak symmetry is obtained as,
\begin{align} 
 \mathcal{M}_D=\begin{pmatrix}
		m_\Psi & M_D \\
		M_D & m_\chi 
		\end{pmatrix},
		\label{eq:massmatrix}
\end{align}
where we define $M_D=\frac{Yv}{\sqrt{2}}$. After diagonalisation of Eq.(\ref{eq:massmatrix}), we are left with two neutral Dirac particles which we identify as $\xi_1$ and $\xi_2$. The mass eigenvalues of $\xi_1$ and $\xi_2$ are given by,
\begin{align}\label{massE}
&m_{\xi_1} \approx m_\chi - \frac{M_D^2}{m_{\Psi}-m_\chi}\\
&m_{\xi_2} \approx m_\Psi + \frac{M_D^2}{m_{\Psi}-m_\chi}
\end{align} 
Therefore, the lightest eigenstate is $\xi_1$, which is the stable DM candidate of our framework. The stability of the DM is ensured by the unbroken $\mathcal{Z}_2$ symmetry. The mixing between two flavor states, {\it i.e.} neutral part of the doublet ($\psi^0$) and the singlet field ($\chi$) is parameterised by,
\begin{align}
\sin2\theta\simeq \frac{2Yv}{\Delta M},
\end{align}
where $\Delta M =m_{\xi_2}-m_{\xi_1}\approx m_\Psi-m_\chi$ in the small $Y$ limit.
Here, $\xi_1$ can be identified with the singlet $\chi$. Since the present analysis involves freeze-in dark matter production, the mixing between the singlet and doublet fermions is always very tiny. Therefore, henceforward we identify the dark matter candidate as $\chi$.

\section{Dark matter relic}\label{DMPH}
The present set-up has two stable particles $\phi$ and $\chi$, and both can contribute to total DM relic abundance. The relative contribution of each component to the DM relic is dependent on early Universe history, as we discuss below. We have worked with $m_\phi\ll m_{\chi}$ and $\lambda\sim\mathcal{O}(10^{-1})$ with an aim to obtain a large amount of self-interaction for $\chi$ that can address the problems associated with small scale structures of the Universe. Irrespective of the cosmological history of the Universe, we assume $a_3,\lambda_{\phi H},\delta, Y \ll 1$ always such that both $\phi$ and $\chi$ can never equilibrate with the SM bath. Additionally, we have considered a specific choice $a_3,\delta,\lambda_{\phi H}\ll Y$ which implies $\phi$ can only be populated at a non-negligible rate (depending on the expansion rate of the Universe) from the $\chi\chi\rightarrow\phi\phi$ process.

\subsection{Radiation dominated Universe}\label{sec:CI}

In the standard description of Big bang cosmology, the Universe is radiation dominated prior to BBN. Initially both the components begin with zero number density and first $\chi$ gets produced from the SM bath. To be specific, production of $\chi$ at early Universe occurs primarily from $\Psi$ decay with thermally averaged decay width $\langle \Gamma_\Psi \rangle^T$ where $T$ is the SM temperature. Before EW symmetry breaking, the DM production takes place only from the $Y\Psi\tilde{H}\chi$ vertex. After EW symmetry breaking, the production channels for DM are namely, $\psi^{\pm}\rightarrow \chi W^{\pm}$, $\xi_2\rightarrow \chi Z$ and $\xi_2\rightarrow \chi h$. In order to express the $\chi$ production rate as function of temperature, we define $\langle\Gamma_\Psi\rangle^T=\langle\Gamma_\Psi\rangle~ \Theta(T-T_{\rm EW})+\langle\Gamma_{\Psi^\pm}+\Gamma_{\xi_2}\rangle~\Theta(T_{\rm EW}-T)$ where $T_{\rm EW}\sim 160$ GeV indicates the EW symmetry breaking temperature. Subsequently, the $\chi\chi\rightarrow \phi\phi$ process yields $\phi$ and when the number density of $\phi$ is sufficient, it can further annihilate to $\chi$ and form a local dark sector thermal equilibrium with an uniform temperature $T_D$ provided
\begin{align}
r \equiv \frac{n_\chi \langle\sigma v\rangle_{\chi\chi\to\phi\phi}}{\mathcal{H}}  \gg 1. 
\label{eq:defr}
\end{align}
Finally $\chi$ freezes out and both contribute non-negligibly to the total relic abundance.
The set of Boltzman equations that governs such dynamics with $z (=\frac{m_{\chi}}{T})$ are given by~\cite{Du:2020avz, Krnjaic:2017tio, Berger:2018xyd, Cheung:2010gj, Cheung:2010gk, Chu:2011be},
\begin{align}
 \frac{d Y_\phi}{d z}=\frac{s}{\mathcal{H} z}\langle \sigma v\rangle_{\chi\chi\rightarrow \phi\phi}^{T_D}\left[Y_{\chi}^2-\left(\frac{Y_\chi^{\rm eq}(T_D)}{Y_\phi^{\rm eq}(T_D)}\right)^2Y_\phi^2\right]\label{eq:rhodBoltzmannCI1}
 \end{align}
 \begin{align}
&\frac{dY_{\chi}}{dz}=-\frac{s}{\mathcal{H} z}\langle \sigma v\rangle_{\chi\chi\rightarrow \phi\phi}^{T_D}\left[Y_{\chi}^2-\left(\frac{Y_\chi^{\rm eq}(T_D)}{Y_\phi^{\rm eq}(T_D)}\right)^2Y_\phi^2\right]+\frac{45 z^2 }{2\pi^2 m_{\chi}^3}\frac{s {\langle\Gamma_{\Psi}\rangle}^T}{\mathcal{H}}\big( Y_\Psi^{\rm eq}(T)-Y_{\chi}\big).
\label{eq:rhodBoltzmannCI2}
\end{align}

\begin{table}[t]
\begin{center}
\begin{tabular}{ | c | c | c | c | c | c | c | c |}
  \hline
  &$\lambda$& $m_\Psi$ & $m_{\chi}$ & $m_\phi$ & $Y$ & $\Omega_{\chi} h^2$& $\Omega_{\phi} h^2$ \\
  \hline 
  RD-I & 0.13  & 1.5 TeV & 20 GeV  & 10 MeV  & $5.40\times 10^{-10}$ & 0.01 & $0.11$  \\ \hline
  RD-II & 0.085  & 1.5 TeV & 10 GeV  & 10 MeV  & $5.45\times 10^{-10}$ & 0.01 & 0.11  \\ \hline
\end{tabular}
\end{center} 
\caption{Two representative benchmark points that describe the reannihilation scenario in the present framework considering radiation dominated Universe.}
\label{tab:Case Ib}
\end{table}    

 To evaluate the dark temperature $(T_{D}$) one needs to solve the Boltzman equation for the total dark sector energy density ($\rho_D=\rho_\chi+\rho_\phi$) where $\rho_i$ indicates the energy density for an individual component~\cite{Du:2020avz, Krnjaic:2017tio, Berger:2018xyd, Chu:2011be, Zhu:2021pad, Hryczuk:2021qtz, Tapadar:2021kgw}. 
\begin{align}
 \frac{d\rho_{D}}{dt}+3 H(\rho_{D}+p_{D})& =\mathcal{P}_{\Psi}n_\Psi^{\rm eq}\label{eq:darkTempCI},
\end{align}
where the quantity $\mathcal{P}_{\Psi}$ represents the thermally
averaged energy transfer rate from the SM to dark sector. The notation $p_D=p_\chi+p_\phi$ stands for the sum of the pressures for individual components. The standard form of $p_D$ is given by,
\begin{align}
 p_D=\frac{g_\chi}{6\pi^2}\int_{m_\chi}^\infty \frac{(E_\chi^2-m_\chi^2)^{3/2}}{e^{E_\chi/T_D}+1} dE_\chi+\frac{g_\chi}{6\pi^2}\int_{m_\phi}^\infty \frac{(E_\phi^2-m_\phi^2)^{3/2}}{e^{E_\phi/T_D}+1} dE_\phi,
 \label{eq:pressure}
\end{align}
The symbols $g_\chi$ and $g_\phi$ imply the internal degrees of freedom for $\chi$ and $\phi$ species. Additionally, one can write the energy density of the dark sector as,
\begin{align}
 \rho_{D}=\frac{g_\chi}{2\pi^2}\int \frac{(E_{\chi}^2-m_{\chi}^2)^{1/2} E_{\chi}^2 dE_{\chi}}{e^{\frac{E_{\chi}}{T_D}}+1}+\frac{g_\phi}{2\pi^2}\int \frac{(E_\phi^2-m_\phi^2)^{1/2} E_\phi^2 dE_\phi}{e^{\frac{E_\phi}{T_D}}+1}.
 \label{eq:TempCI} 
\end{align} 
Analytical computation of the integrations in RHS of Eqs.(\ref{eq:pressure}-\ref{eq:TempCI}) in presence of Fermi-Dirac distribution function looks complicated. Instead one can compute the same in the Maxwell Boltzman-approximation. We find  \footnote{We verified a close match between the numerical estimation from the standard expression with this approximation for our calculation.},
\begin{align}
 & p_D=\frac{g_\chi }{2\pi^2}m_\chi^2 T_D^2K_2\left(\frac{m_\chi}{T_D}\right)+ \frac{g_\phi}{2\pi^2}m_\phi^2 T_D^2K_2\left(\frac{m_\phi}{T_D}\right)\label{eq:rhoD}\\
 & \rho_D=\frac{g_\chi m_\chi^2}{2\pi^2}T_D\left[m_\chi K_1\left(\frac{m_\chi}{T_D}\right)+3 T_D K_2\left(\frac{m_\chi}{T_D}\right)\right]+\frac{g_\phi m_\phi^2}{2\pi^2}T_D\left[m_\phi K_1\left(\frac{m_\phi}{T_D}\right)+3 T_D K_2\left(\frac{m_\phi}{T_D}\right)\right]\label{eq:pD}.
\end{align}
Replacing Eq.(\ref{eq:rhoD}) and Eq.(\ref{eq:pD}) in Eq.( \ref{eq:darkTempCI}), it is simple obtain the evolution equation for $T_D$ as function of SM temperature.

\begin{figure}[t]
 \includegraphics[height=6cm,width=7.4cm]{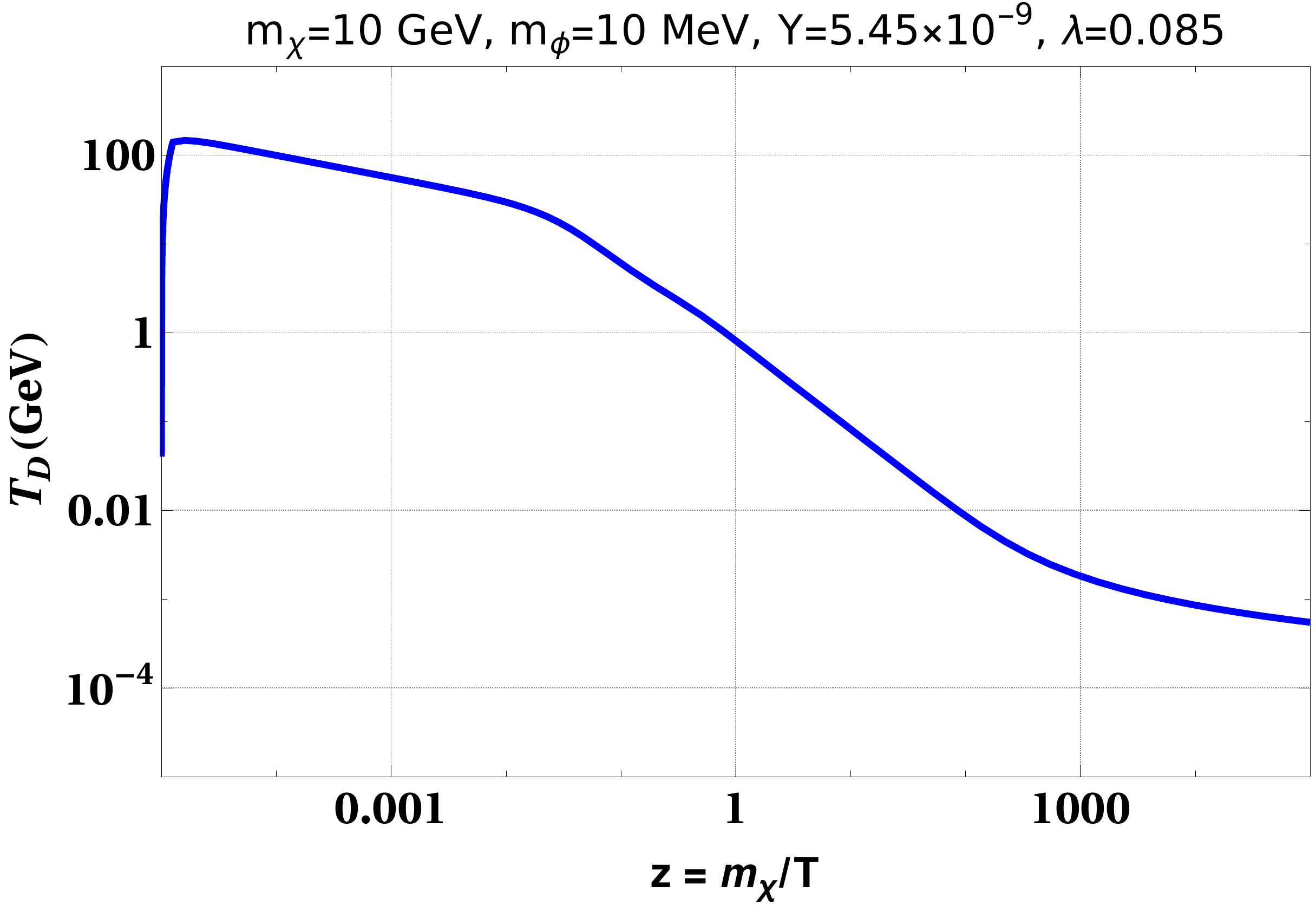}
 ~~\includegraphics[height=6cm,width=7.4cm]{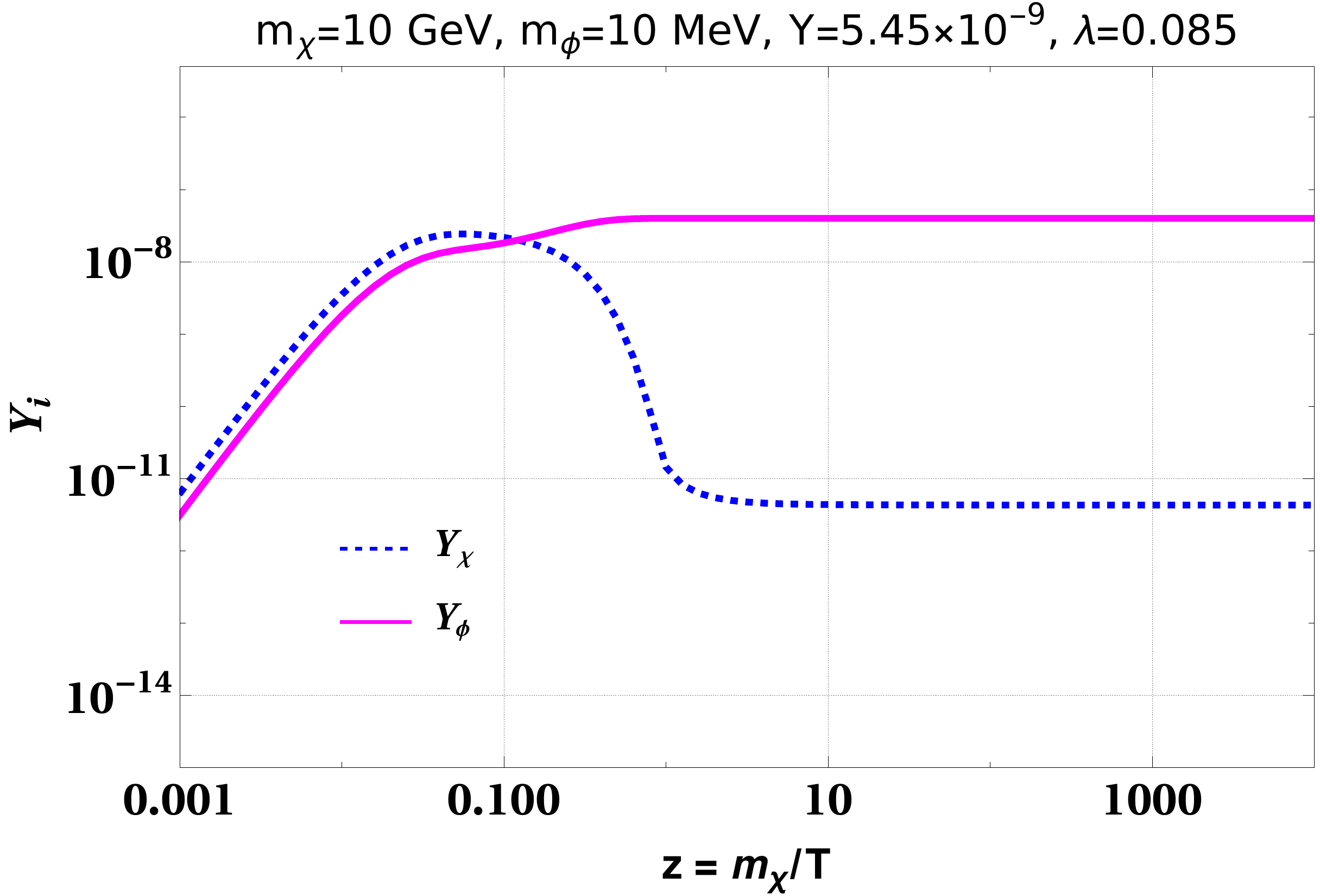}
 \caption{The left plot shows the variation of dark sector temperature $T_D$ as a function of inverse temperature $z$. Corresponding evolutions for the number densities of dark sector particles are shown in the right plot. Here, after initial non-thermal production of $\chi$ strong self-interaction generates a dark sector thermal equilibrium. Further, freeze-out produces the $\phi$ abundance. The figures are computed considering the benchmark point RD-II in Table \ref{tab:Case Ib}.}
 \label{fig:CaseIb-I}
\end{figure} 

\begin{figure}[t]
\begin{center}
\includegraphics[height=7cm,width=9.0cm]{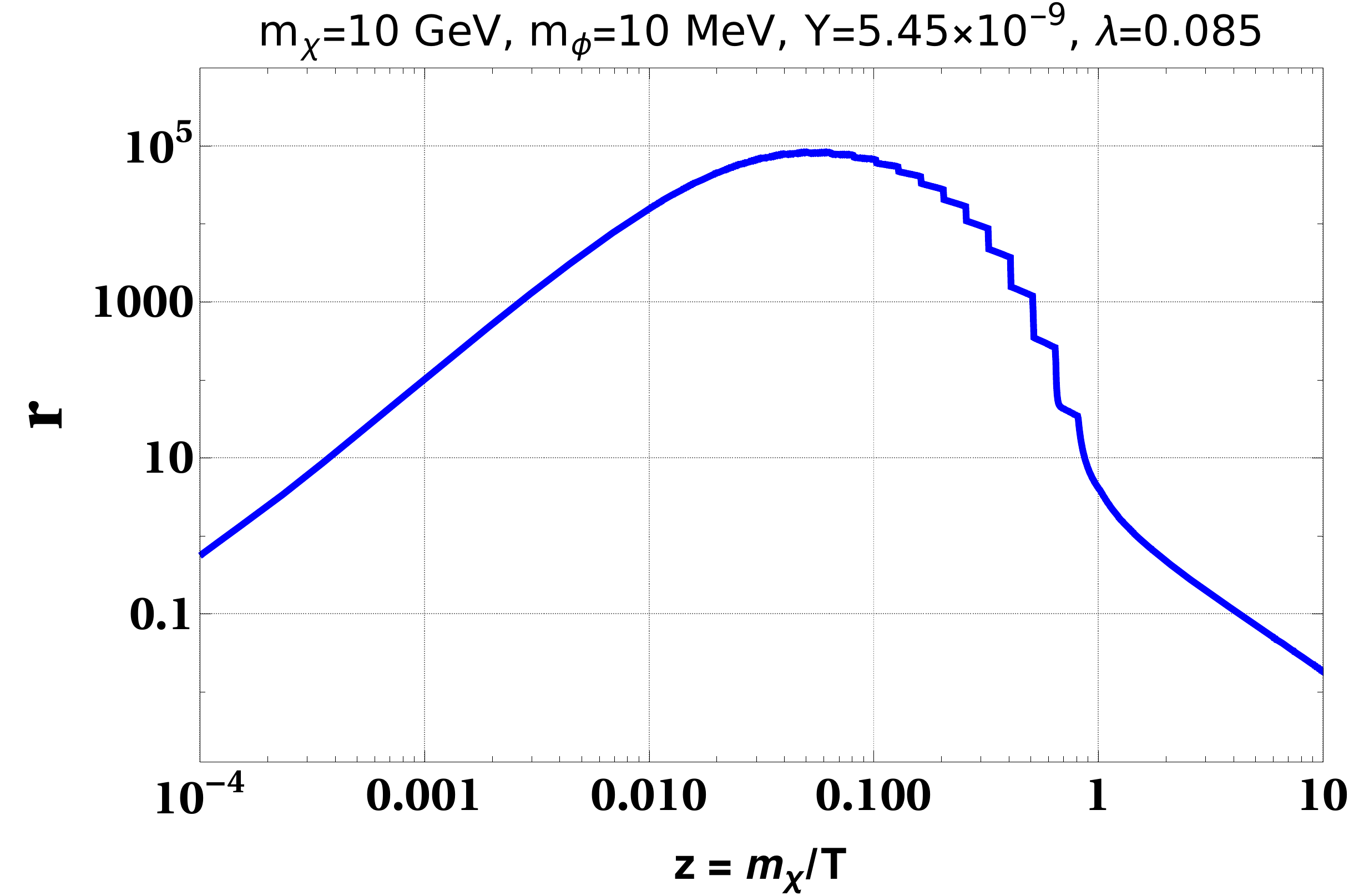}
\end{center}
 \caption{The ratio of the interaction rate among the dark sector particles and Hubble parameter of the Universe following Eq.(\ref{eq:defr}) as function of temperature for for the benchmark point RD-II in Table \ref{tab:Case Ib}.}
 \label{fig:CaseIb-II}
\end{figure} 
In Table \ref{tab:Case Ib}, we provide two reference points that instigate dark freeze out of $\chi$ after the non-thermal production from the SM bath for two choices of dark matter masses having magnitude 10 GeV and 20 GeV, respectively.
The evolutions of dark sector temperature and $\chi$ and $\phi$ abundances for the benchmark point RD-II (with DM mass 10 GeV) are shown in Figure \ref{fig:CaseIb-I}. To generate these figures, we solve the set of coupled Boltzmann equations as described in Eqs.(\ref{eq:rhodBoltzmannCI2}-\ref{eq:TempCI}) and also used Eqs.(\ref{eq:rhoD}-\ref{eq:pD}) with the initial conditions $(Y_{\chi},~Y_\phi,~T_D)=0$ and $Y_\Psi=Y_\Psi^{\rm eq}(T)$. In Figure \ref{fig:CaseIb-II}, we display the ratio $r$ as defined in Eq.(\ref{eq:defr}) to measure the strength of dark sector interaction rate as a function of SM temperature for the same benchmark point. This figure confirms the formation of dark sector equilibrium after the non-thermal production of both the components with $r\gg 1$ for a finite period. In the left of Figure \,\ref{fig:CaseIb-I}, we see the evolution pattern of the dark temperature as a function of SM bath temperature. Since the dark matter annihilation to SM particles is negligible, it turns out that $T_D\ll T$ always.
 Initially, the dark temperature increases from zero due to constant entropy injection from the SM bath and reaches a maximum value. Then, due to the expansion of the Universe, it keeps decreasing. Till $T\sim m_\Psi$, the production of $\chi$ continues from $\Psi$ decay, and hence we see a slower decreasing rate of $T_D$ initially, and after that, it becomes steep. 
 
 The most intriguing part is, here, due to large $\lambda$, the $\chi$ keeps annihilating to $\phi$ till the decoupling and ends with suppressed abundance (see right of Figure \,\ref{fig:CaseIb-I}). Hence for the chosen benchmark points in the reannihilating scenario, we notice $\Omega_{\chi}\ll \Omega_{\phi}$. An enhancement of $\lambda$ for a fixed $m_\chi$ may slightly increase the relic of $\phi$ further, while it will be reduced further in the opposite limit. However, a small $\lambda$ is not desirable in view of generating sufficient self-interaction of $\chi$ to solve the small scale structure problems of the Universe. It is to note that for both the reference points in Table \ref{tab:Case Ib}, the total relic abundance $\Omega_T h^2=\Omega_\phi h^2+\Omega_{\chi}h^2\sim 0.12$ \cite{Planck:2018vyg} remains within the observed limit by Planck requiring a similar order of $Y$ value.

Thus, radiation dominated Universe with a large $\lambda$ fails to provide a pure $\chi$ dominated scenario due to the late time annihilation of the dark matter $\chi$ to mediator $\phi$.
We anticipate that a non-standard Universe in the form of kination or faster than kination domination could enforce significant suppression of $\phi$ production rate from $\chi\chi\rightarrow \phi\phi$ process even in the presence of large $\lambda$. This may occur if the ratio in Eq.(\ref{eq:defr}) remains comparatively suppressed till late time. Note that such faster expansion of the Universe may also slow down the production rate of $\chi$ from the SM bath. However, that can be increased by adjusting $Y$.

\subsection{Non-standard Universe}
In standard description of cosmology, we generally assume that the Universe is radiation dominated in the pre big bang nucleosynthesis (BBN) era. However, due to lack of evidences,  possibilities remain open that before BBN, the Universe could have been occupied by a nonstandard fluid, redshifting faster or slower than the radiation component. Earlier works have emphasised that consideration of such modified cosmology poses non-trivial impact on the dark matter phenomenlogy.

A non-standard Universe can be sketched by assuming the presence of an additional species ($\eta$) along with the radiation component in the Universe. We consider the equation of parameter ($\omega$) of the nonstandard fluid is larger than that of radiation. We parameterize this by $\rho_\eta\propto a^{-3(1+\omega)}$ which can be converted to $\rho_\eta\propto a^{-(4+n)}$ with $\omega=\frac{1}{3}(n+1)$ and $n>0$. The modified description of the Universe leads to the redefinition of the Hubble parameter ($\mathcal{H}$) as given by~\cite{DEramo:2017gpl, Konar:2020vuu, Barman:2021ifu},
 \begin{align}
 \mathcal{H}^2=\frac{\rho_R+\rho_\eta}{3 M_P^2},
 \end{align}
where $M_P$ refers the reduced Planck scale, $\rho_R$ and $\rho_\eta$ correspond to the energy densities of radiation and $\eta$ component. The total energy density of the Universe in presence of $\eta$ as function of temperature can be expressed as, 
\begin{align}\label{totalrho}
 \rho(T) &= \rho_{\rm rad}(T)+\rho_{\eta}(T)\\
 &=\rho_{\rm rad}(T)\left[1+\frac{g_* (T_r)}{g_* (T)}\left(\frac{g_{*s}(T)}{g_{*s}(T_r)}\right)^{(4+n)/3}\left(\frac{T}{T_r}\right)^n\right],
\end{align}
where $\rho_R=\frac{\pi^2}{30}g_*(T)T^4$ with $g_*$ stands for the number of relativistic degrees of freedom. The relativistic entropy degrees of freedom is denoted by $g_{*s}$. Thus a faster expanding Universe is simply parameterised by the set of $(n,T_r)$. A larger $n$ or smaller $T_r$ prompts the energy density of the Universe to redshift more faster. The temperature $T_r$ implies the end of modified expansion rate and we get back the radiation dominated phase. In case of standard radiation dominated Universe, $\eta$ field would be absent and we simply consider $\rho=\rho_{\rm rad}$.  
The BBN observation on the number of relativistic degrees of freedom imposes a lower bound on $T_r~(\gtrsim (15.4)^{1/n}$ MeV). A special case $n=2$ (or $\omega=1)$ is familiar as kination domination phase. For $n>2$, one has to consider
scenarios faster than quintessence with negative potential. We refer the readers to \cite{DEramo:2017gpl}, for a detailed description of a fast-expanding Universe. Below we separately discuss the cases for kination domination and faster than kination domination.

 \begin{table}[t]
\begin{center}
\begin{tabular}{ | c | c | c | c | c | c | c | c | c |}
  \hline
  & $\lambda$ & $m_\Psi$ & $m_{\chi}$ & $m_\phi$ & $Y$ & $n$ & $T_r$ &  $\Omega_{\chi} h^2$ \\
  \hline 
  FKD-I & 0.13  & 1.5 TeV & 20 GeV  & 10 MeV  & $2\times 10^{-8}$ & 3 & 20 MeV & 0.119  \\ \hline
    FKD-II & 0.085  & 1.5 TeV & 10 GeV  & 10 MeV  & $2.44\times 10^{-8}$ & 3 & 22 MeV & 0.119  \\
  \hline
\end{tabular} 
\end{center}
\caption{Two representative benchmark points that demonstrate \textit{pure freeze-in} non-standard cosmology with faster than kination domination.}
\label{tab:CIIa}   
\end{table}    

\subsubsection{Faster than kination phase}\label{sec:CIIa}
As earlier mentioned a faster than kination phase is realised for $n>2$. Here we consider $n=3$. A faster expansion implies enhanced Hubble rate which can suppress
the interaction rate among the dark sector particles. If the interaction rate among the dark sector particles gets heavily suppressed, that leads to the production of $\chi$ by pure freeze-in. In that case the governing Boltzman equation for $\chi$ looks very simple as given by~\cite{Hall:2009bx, Konar:2021oye},
\begin{align}
 \frac{dY_{\chi}}{dz}=\frac{45 z^2 }{2\pi^2 m_{\chi}^3}\frac{s {\langle\Gamma_{\Psi}\rangle}^T}{\mathcal{H}}\big( Y_\Psi^{\rm eq}(T)-Y_{\chi}\big).
\end{align}
Since the rate of the conversion process $\chi\chi\to\phi\phi$ is negligibly low in case of pure freeze-in, we do not estimate the relic of $\phi$ here, which is expected to be barely abundant in the present Universe. For the same reason, the computation of dark temperature is not essential here.

\begin{figure}[h]
 \includegraphics[height=6.4cm,width=7.4cm]{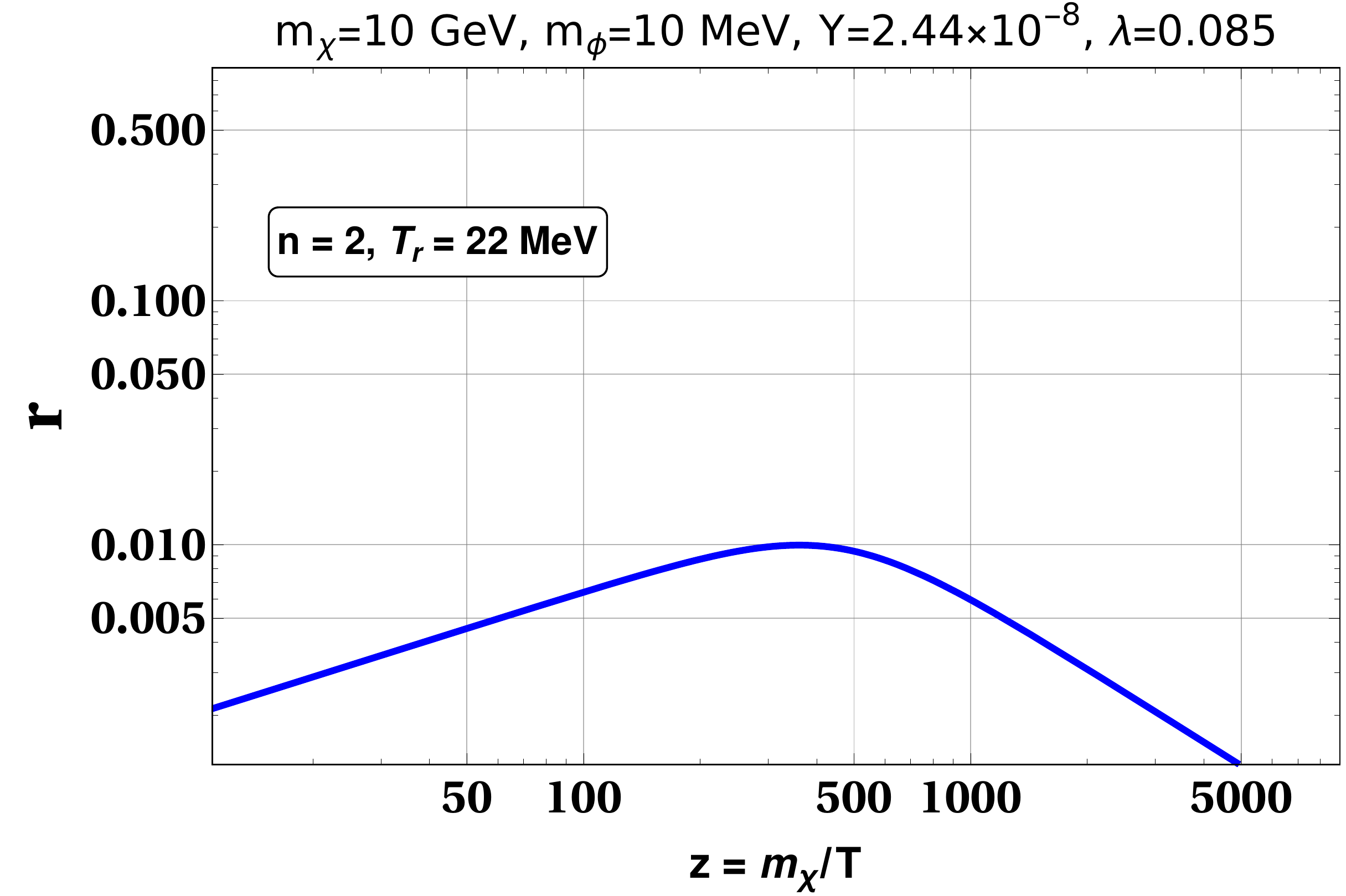}
 ~~\includegraphics[height=6.4cm,width=7.4cm]{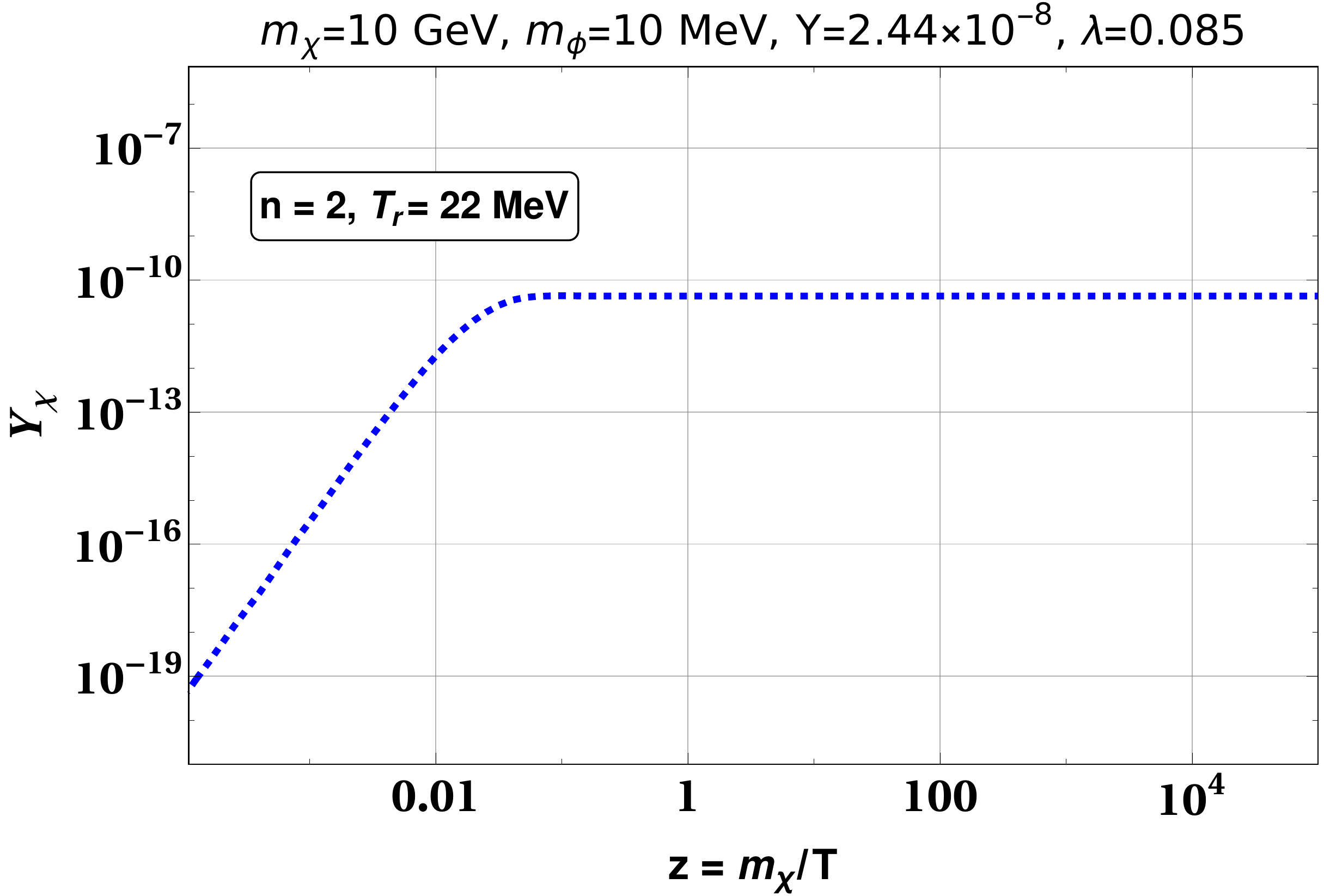}
 \caption{Left: The ratio of the interaction rate among the dark sector particles and Hubble paramter of the Universe as function of temperature for the benchmark point FKD-II in Table \ref{tab:CIIa}. Right: The evolution of DM comoving number density is shown with temperature for the same reference point.}
\label{fig:CIIa}  
 \end{figure} 

\begin{figure}[t]
 \begin{center}
 \includegraphics[height=8cm,width=11cm]{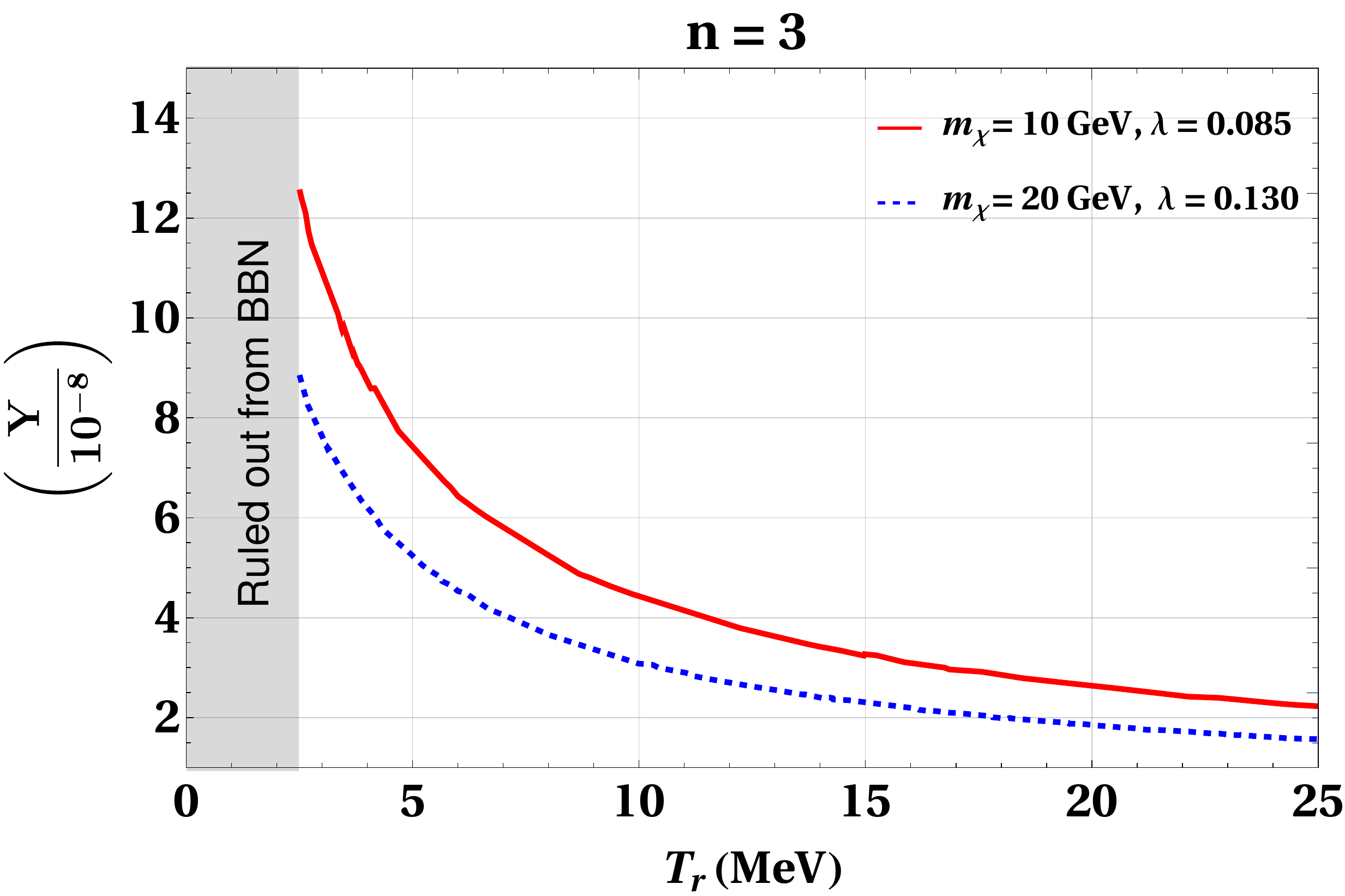}
 \end{center}
 \caption{Relic satisfied contours in $T_r-Y$ plane considering $n=3$ for $m_{\chi}=10$ GeV and 20 GeV.}
 \label{fig:YvsTr}
 \end{figure}

In Table \ref{tab:CIIa}, we note down two benchmark points with BSM model inputs $\{\lambda,m_\Psi,m_\chi,m_\phi\}$ are same as used in section \ref{sec:CI}. We have considered zero initial abundance of $\chi$ since it remains out of equilibrium in the early Universe. We fix $n=3$ for this case and vary the $T_r$ and $Y$ in order to obtain correct relic abundance (by pure freeze-in) as allowed by Planck. Recall that these two reference points have portrayed reannihilation patterns with $\Omega_{\chi} h^2\ll \Omega_\phi h^2$ for a radiation dominated Universe. 
 In the right of Figure \,\ref{fig:CIIa}, we have depicted the evolution of comoving abundance of $\chi$ with temperature considering the reference point FKD-II.
 We have used $n=3$ and fixed the parameters $T_r$ and $Y$ such that correct relic abundance for $\chi$ is attained. In left of Figure \,\ref{fig:CIIa}, we estimate the parameter $r$ as earlier defined and found it to be order of $\sim \mathcal{O}(10^{-2})$. This indeed ensures the validity of labelling the present scenario as \textit{pure freeze-in}. Here we do not mention the abundance of the $\phi$ field. Owing to the faster expansion of the Universe, the produced $\chi$ never creates its own bath, and it is expected that the contribution of $\chi\chi\rightarrow\phi\phi$ would be negligible in yielding $\phi$.
 Nevertheless, a conservative test can always be performed by solving the Boltzmann equations for $Y_\phi$ and $Y_{\chi}$ while considering the maximum~\footnote{Maximum possible value of a thermally averaged cross section can be found by equating the temperature of the corresponding bath with the heavier mass scale approximately associated with the interaction.} possible value of $\langle\sigma v\rangle_{\chi\chi_\rightarrow \phi\phi}$. By utilising this conservative method, we have found similar results (maximum uncertainty is 2$\%$) with $Y_\phi$ remaining much lower than $Y_{\chi}$. We also notice from Table \ref{tab:CIIa}, that the required order of the Yukawa coupling is relatively larger than the one shown in section\,\ref{sec:CI} considering standard RD Universe. The reason is obvious as a faster expanding Universe not only suppresses the rate of $\chi\chi\rightarrow \phi\phi$ process but also slows down the production process of $\chi$ and therefore, one needs to raise BSM Yukawa coupling parameter $Y$ appropriately. To investigate the $T_r$ dependence on $Y$ further, in Figure \,\ref{fig:YvsTr}, we show the relic density satisfied contours in the $T_r-Y$ plane for $m_\chi=10$ GeV and 20 GeV considering $n=3$. We keep the corresponding values of $\lambda$ and $m_\Psi$ in accordance with Table\,\ref{tab:CIIa}. For a fixed DM mass, a smaller $T_r$ requires larger $Y$. This occurs since a smaller $T_r$ implies an enhanced Hubble rate, and it requires a larger $Y$ to obey the relic abundance bound.
 Moreover, a smaller DM mass also requires larger $Y$ to obey the relic bound. It is to note that in Figure \,\ref{fig:YvsTr}, we have kept $T_r$ below 25 MeV. Beyond this value of $T_r$, the ratio $r$ would turn larger than $\mathcal{O}(10)^{-2}$ and therefore, the contribution of $\chi\chi\to\phi\phi$ conversion in the final DM relic may turn important at some extent.

 In summary, the above discussion reveals that a faster than kination dominated early Universe ($n=3$) is able to provide a pure freeze-in yield for $\chi$ with absolutely dominant share to the total relic abundance even in the presence of $\lambda\sim \mathcal{O}(10^{-1})$. This is in sharp contrast to the RD Universe where we found $\Omega_\chi h^2\ll \Omega_\phi h^2$ with the same values of $\lambda$ and mass scales of the dark sector particles.
 In the upcoming section, we discuss how the scenario evolves in the case of kination dominated ($n=2$) early Universe. 
 
\subsubsection{Kination phase}\label{sec:CIIB} 
\begin{figure}[t]
 \begin{center}
 \includegraphics[height=8cm,width=11cm]{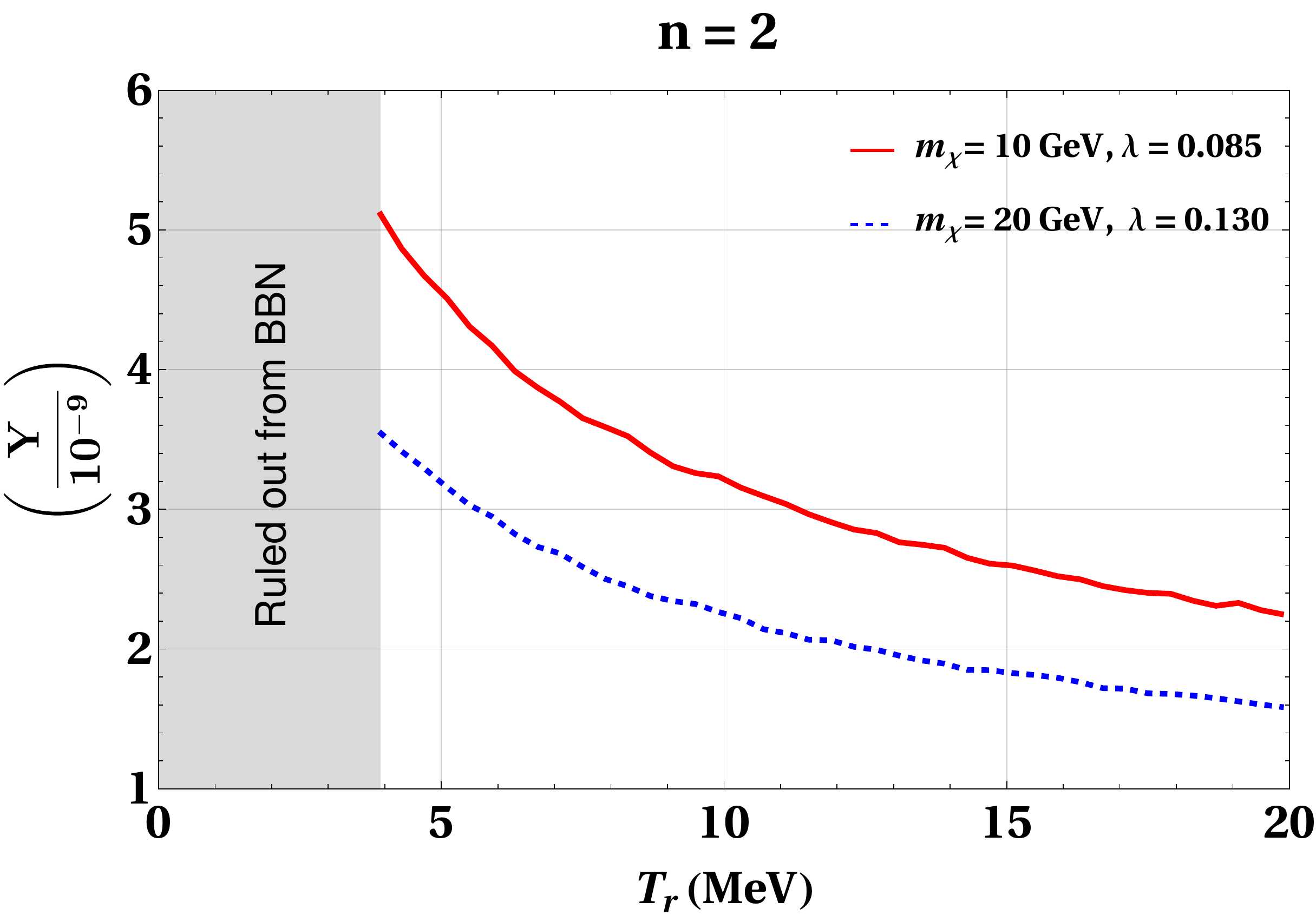}
 \end{center}
 \caption{Relic satisfied contours in $T_r-Y$ plane for $m_{\chi}=10$ GeV and 20 GeV considering $n=2$.}
 \label{fig:YvsTr1}
 \end{figure}

Earlier, we have seen, faster than kination domination has led to pure freeze-in for the parameters that show reannihilation in the case of an RD Universe by suppressing the interaction rate among dark sector particles. Now, we consider the kination domination case ($n=2$), which leads to a relatively slower expansion rate of the Universe than $n=3$ and hence may enhance the interaction rate inside the dark sector. Thus to realise pure freeze-in with kination domination era, a relatively smaller value of $Y$ for a fixed $T_r$ is expected to obey the relic bound compared to the $n=3$ case. Indeed such prediction emerges to be correct as evident from Figure \,\ref{fig:YvsTr1}. Using the approximate method as commented earlier, we have found that for the respective ranges for $Y$ and $T_r$ in 
 Figure \,\ref{fig:YvsTr1}, the ratio $r$ always remains $\lesssim \mathcal{O}(10^{-2})$ and thus relic $\phi$ is always suppressed with an uncertainity $5\%$ atmost.
 
One may also wonder whether a reannihilating scenario for the dark matter $\chi$ with final correct relic abundance is possible in non-standard cosmological models. To investigate this, let us stick to the $n=2$ case. This would obviously require a larger $T_r$ compared to the pure freeze-in case with $n=2$.
Assuming internal equilibrium of dark sector is established for some period in the early Universe, the set of Boltzman equations remain same as in Eq.(\ref{eq:rhodBoltzmannCI1}-\ref{eq:darkTempCI}).
We would like to reuse the same set of benchmark values for ($\lambda,m_\Psi,m_\chi,m_\phi$) as originally introduced in section \ref{sec:CI}.
We would like to emphasise that, these points depict a reannihilation pattern in the radiation dominated Universe, while the presence of a kination or faster than kination dominated epoch transform it to pure freeze-in with almost full $\chi$ occupancy in the total relic abundance provided suitable choices for $T_r$ are made. Here, we find out the estimates of $(Y, T_r)$ such that it shows a reannihilation pattern with $\Omega_\chi h^2\gg \Omega_\phi h^2$.
We consider zero initial abundance for the dark sector particles as well as $T_{D}^{\rm ini}=0$. In Table \ref{tab:CIIb}, we note down the required order of $T_r$ and the Yukawa coupling that predicts such scenario considering the same set of ($\lambda, m_\Psi, m_\chi, m_\phi$) as used in section \ref{sec:CI}. 

\begin{table}[t]
\begin{center}
\begin{tabular}{ | c | c | c | c | c | c | c | c | c | c|}
  \hline
  & $\lambda$ & $m_\Psi$ & $m_{\chi}$ & $m_\phi$ & $Y$ & $n$ & $T_r$& $\Omega_{\rm \chi} h^2$ & $\Omega_{\phi} h^2$ \\ 
  \hline  
  KD-I & 0.13  & 1.5 TeV & 20 GeV  & 10 MeV  & $2.4\times 10^{-9}$ & 2 & 320 MeV & 0.115 & 0.003  \\
  \hline 
  KD-II & 0.085  & 1.5 TeV & 10 GeV  & 10 MeV  & $3.75\times 10^{-9}$ & 2 & 180 MeV & 0.118 & 0.001  \\
  \hline
\end{tabular}    
\end{center}
\caption{The representative benchmark points that lead to reannhilation after freeze-in production of $\chi$ in presence of kination dominated epoch with $n=2$.}   
\label{tab:CIIb}
\end{table}    

\begin{figure}[t]
 \includegraphics[height=6.6cm,width=7.4cm]{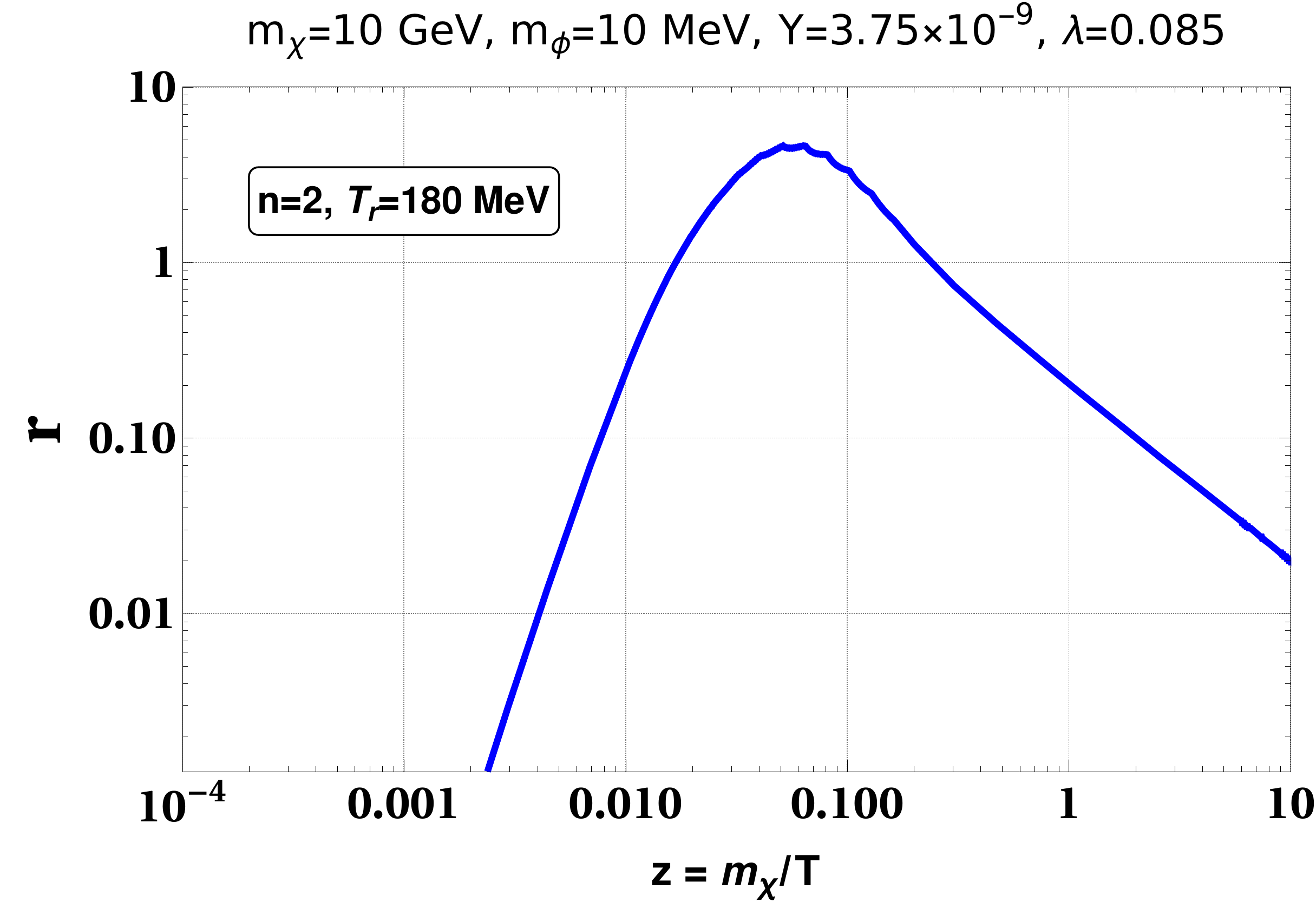}
 ~~\includegraphics[height=6.4cm,width=7.4cm]{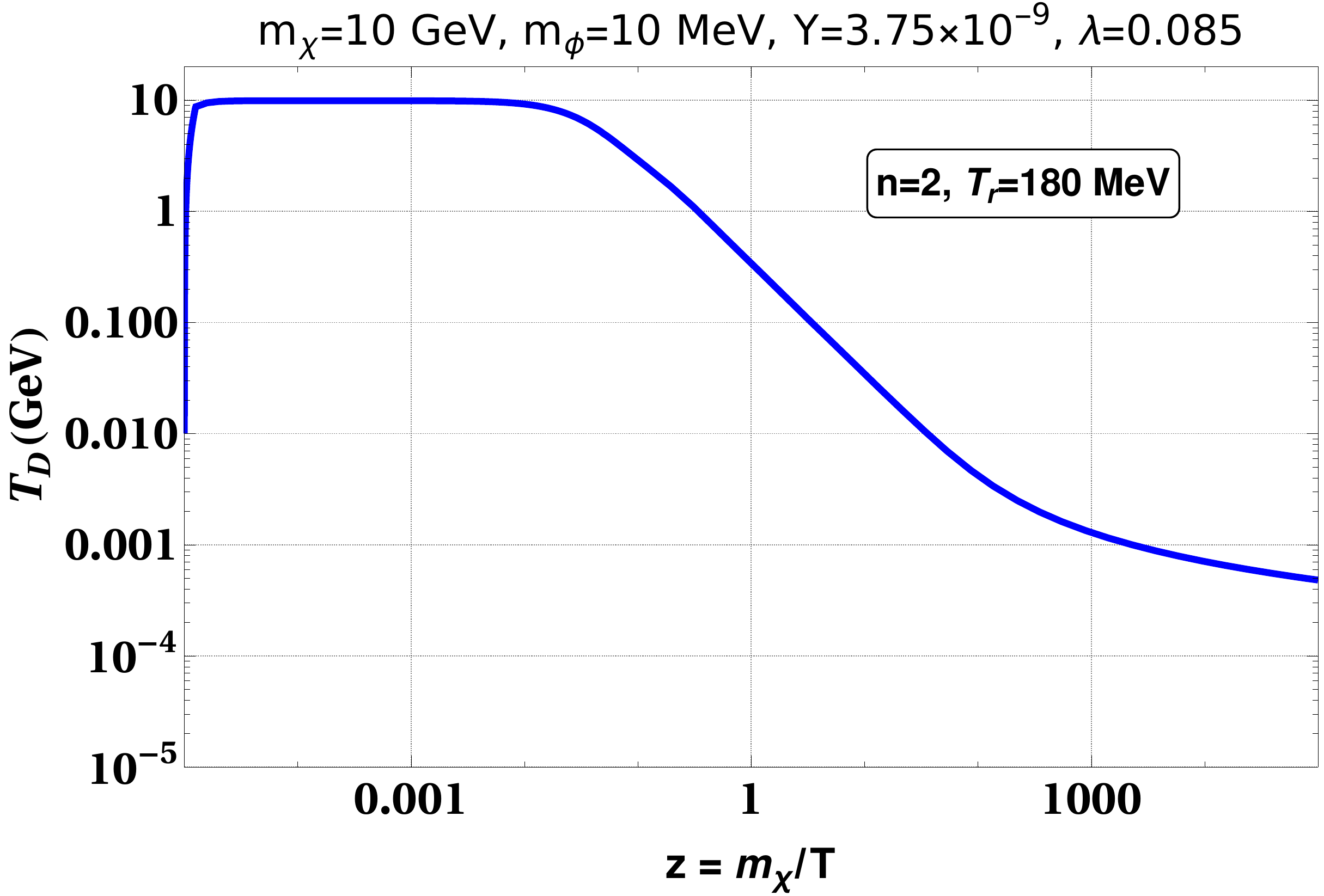}
 \caption{Left: The variation of parameter $r$ is shown against the SM temperature for the reference point KD-II. Right: The variation of $T_D$ on SM temperature $T$ is shown for the same reference point.}
 \label{fig:CIIb} 
\end{figure}

\begin{figure}[t]
 \begin{center}
 \includegraphics[height=7.1cm,width=9.0cm]{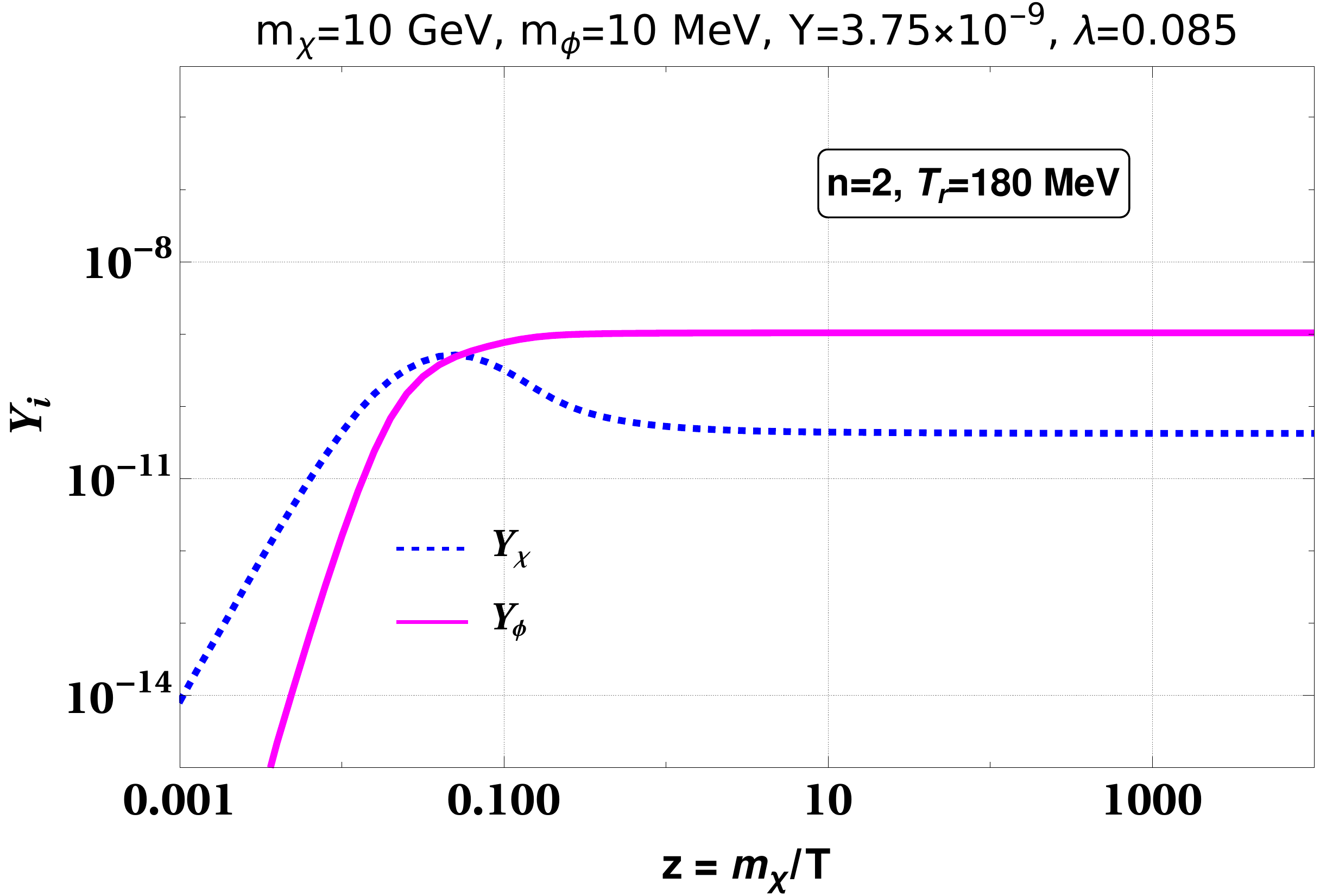}
 \end{center}
 \caption{The evolution of $\chi$ and $\phi$ number densities are shown with SM temperature for the benchmark point KD-II.} 
 \label{fig:KD-Y}
\end{figure}

The evolution patterns of $Y_\chi$ and $Y_\phi$ for the benchmark point KD-II in Table \ref{tab:CIIb} are similar to the case of RD Universe as shown in Figure \ref{fig:CIIb}. In the left of Figure \ref{fig:CIIb}, the ratio $r$ is plotted as a function of the temperature of the SM bath. This figure shows $r$ crosses unity for a brief period and dark sector equilibrates. This feature enables $\chi$ to annihilate at a late time to $\phi$ after production from the standard model bath. However, the annihilation rate is suppressed due to faster expansion of the Universe, which results in reduced relic for $\phi$ as compared to the one in RD Universe. We also notice such an enhanced expansion rate of the Hubble parameter slows down the $\chi$ production process itself, but that can be adjusted by tuning the Yukawa coupling $Y$ appropriately (see Table \ref{tab:CIIb}). 
In right of Figure \ref{fig:CIIb},   
 the temperature $T_D$ is plotted as a function of SM temperature $T$. We see a nontrivial pattern for the temperature evolution of the dark sector here. The $T_D$ remains constant up to $z\sim0.01$ then reduces with $z$. This has occurred specifically due to an accidental cancellation between the second term of LHS in Eq.(\ref{eq:darkTempCI}) and the term in RHS while solving the Boltzman equation for dark sector energy density. Such cancellation is triggered by the same temperature dependence of the Hubble parameter and $n_\Psi^{\rm eq}$ when $\Psi$ is relativistic. {In Figure \ref{fig:KD-Y} the evolution patterns for the number densities of
$\chi$ and $\phi$ considering the benchmark point KD-II are shown as function of SM temperature. The obtained patterns are similar as in
case of RD Universe. The comoving number density of $\chi$ drops from its freeze-in value due to late time
dark freeze out}. A similar reannihilation pattern for $m_\chi=20$ GeV can be obtained as well with the proper assignment of other relevant parameters as pointed out in Table \ref{tab:CIIb}. For both the benchmark points of Table \ref{tab:CIIb}, $\chi$ occupies the maximum share of total relic abundance, whereas $\phi$ can contribute up to 2$\%$ of the total relic.

Before we close this section, let us draw a clear comparison between the impacts of standard and non-standard cosmology within the present setup. We begin by fixing the mediator mass $m_\phi=10$ MeV and dark matter at 10 GeV and 20 GeV, respectively. 
In an RD Universe, a larger value of $\lambda$ (motivated from generating velocity dependent large self interaction) leads to the annihilation of $\chi$ to $\phi$ after its production by forming dark sector equilibrium.
Consequently, we obtain $\phi$ as a dominant DM component rather than $\chi$. We then anticipate that presence of a non-standard epoch (kination or faster than kination) in the early Universe can assist in realising a dominant share of $\chi$ in total DM relic abundance with or without reaching dark sector thermal equilibrium for the same dark sector parameters as used in the RD scenario. We have utilised the same benchmark points as used in the RD case and show that, indeed, a fast-expanding Universe changes the DM dynamics completely and helps in realising $\chi$ as the main dark matter component. This occurs since the presence of a modified cosmology enhance the expansion rate of the Universe and suppresses the conversion rate of the $\chi\chi\to\phi\phi$ process.
One can also spot that it requires a larger amount of Yukawa coupling $Y$ to obey the relic density bound in a non-standard era compared to the case in RD Universe. This, in turn, improves the collider search prospects of the present set up as we will talk about shortly.

\section{Discussion on Collider searches}\label{CC}
We shall now briefly discuss the detection prospects of the proposed singlet scalar extended singlet doublet freeze-in DM at colliders. 
The possible collider signatures of the singlet doublet fermion DM model have been discussed in detail in the context of WIMP \cite{2019,Barman:2019tuo} and FIMP \cite{No:2019gvl, Calibbi:2018fqf} DM scenarios. 
The collider sensitivity of the singlet doublet dark matter model is mainly based on the $Y\overline{\Psi}\tilde{H}\chi$ vertex. Note that the same vertex determines the production efficiency of the dark matter in the early Universe. Depending on the cosmological history and dark sector dynamics, we found that the value of $Y$ keeps changing for fixed DM mass and $m_\Psi$ or $m_{\xi_2}$. For example, a fast expanding Universe prefers larger $Y$ as compared to the RD Universe in each of the pure freeze-in or reannihilation scenarios to obtain a constant dark matter relic.
Therefore in the present framework, the colliders can perhaps be utilised to test the cosmological history at the early Universe and the hidden dark sector dynamics in the context of the singlet doublet model.

\begin{figure}[t]
\begin{center}
 \includegraphics[height=7.5cm,width=11.0cm]{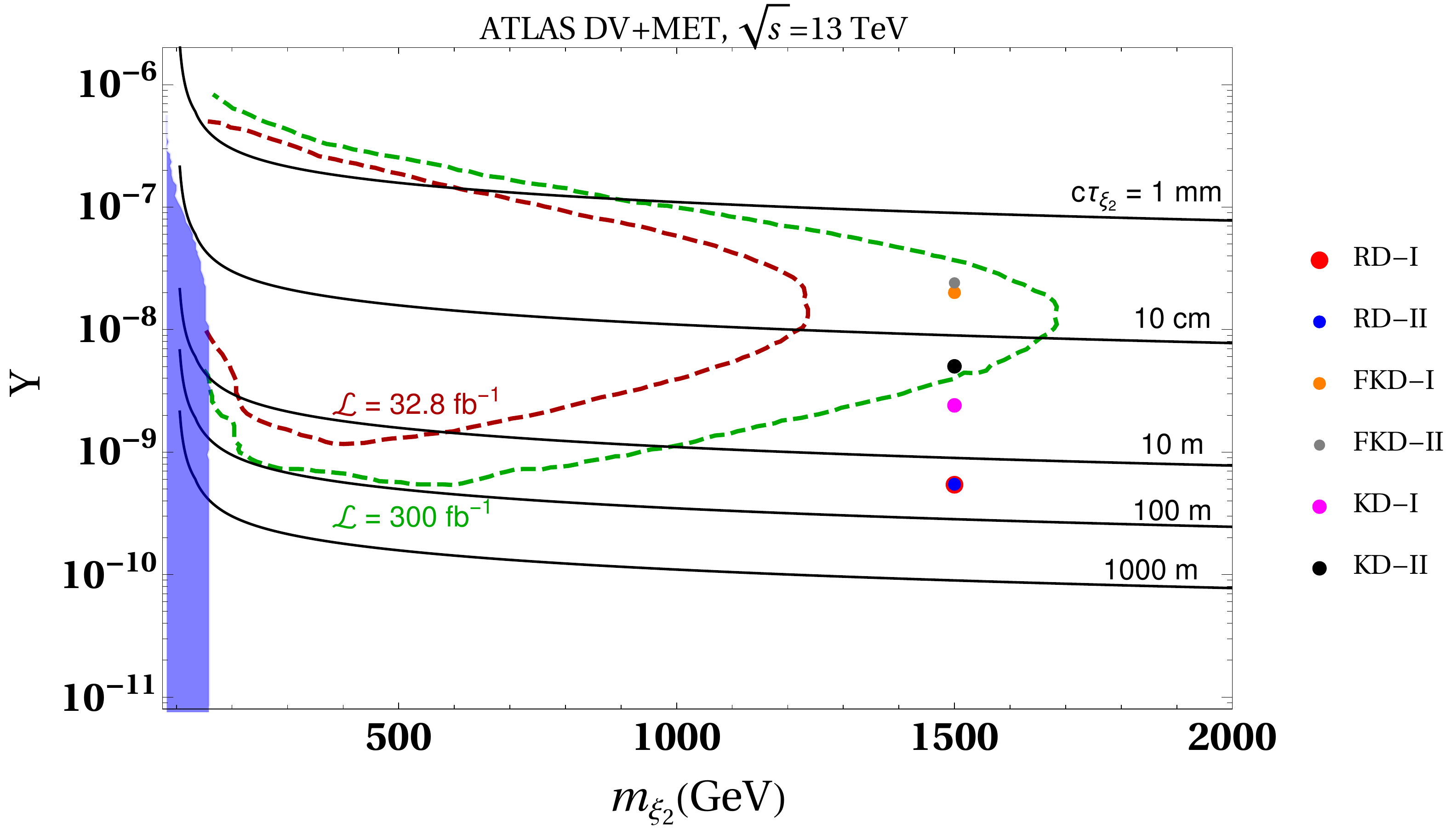}
 \end{center}
 \caption{Discovery prospects of the chosen benchmark points (both in standard and non standard cosmology) in $Y-m_{\xi_2}$ plane. The blue region is disfavored by the search of disappearing charge track signature. We have also shown the present and proposed LHC exclusion sensitivities by the dashed red and green contours respectively as taken from \cite{Calibbi:2018fqf}.}
 \label{fig:llp}
\end{figure}

One of the possible signatures at LHC could be the disappearing charge track signature induced by $\psi^\pm\rightarrow \pi^\pm\xi_2$ decay with decay length $c\tau_{\psi^+}\sim \mathcal{O}(1)$ cm. Independent analyses by ATLAS and CMS collaborations~\cite{ATLAS:2017oal, CMS:2018rea} have inferred strong restriction on the chargino mass considering a supersymmetric framework as a benchmark model. Since the decay $\psi^+\rightarrow \pi^\pm\xi_2$ is equivalent to chargino to Higgsino production, in the present analysis, the bounds provided by ATLAS and CMS can be employed in our analysis as well. In Figure  \ref{fig:llp}, the purple shaded region is disfavored due to non-observation of disappearing charge track assuming $\text{Br}(\psi^{\pm}\to\pi^{\rm}\xi_{2})\simeq 1$. In our analysis, we have considered $m_\Psi$ to be of the TeV scale. Thus the bound arising from the disappearing charge track signature is not important for our case.

On the other hand, due to gauge mediated interactions, the heavy charged ($\psi^\pm$) and neutral (approximately $\xi_2$) components of dark fermion doublet in this model 
can be produced at a hadronic collider. The relevant processes are $p~ p \to \psi^+~\psi^-$, $\xi_2 ~\xi_2$ and $\psi^\pm~ \xi_2$ which further decay to DM ($\chi ~{\rm or}~\xi_1$) along with jets in final states via $hh$, $hZ$, $ZZ$, $W^+ W^-$ modes. The presence of feeble interaction between heavy doublet and the DM ($Y \overline{\Psi} \tilde{H} \chi$) which is the one of requirements of freeze-in scenario makes the heavy neutral state $\xi_2$ longlived ($c\tau_{\xi_2} > 1$ mm) at the typical scale of detector length. The heavy charged state, $\psi^\pm$ promptly decays to a heavier neutral state, $\xi_2$ with a soft pion which is difficult to probe. The charge fermion, $\psi^\pm$ can also decay directly into DM with $W$ ($\psi^\pm \to W^\pm \chi$), which is suppressed when the singlet doublet mixing is small. In the limit of small singlet doublet mixing, the direct and associate productions of $\xi_2~\xi_2$ are one of the promising modes for our analysis which can give rise to displaced vertices with jets plus missing energy signature (DV+MET) before reaching the end of the tracker. Similar kind of search has been performed by ATLAS with $\sqrt{s}=13$ TeV and $\mathcal{L}=32.8$ fb$^{-1}$ in the context of split supersymmetric models (i.e. gluino-neutralino) \cite{ATLAS:2017tny}. The displaced events will be analysed by looking at the individual jet tracks which are originating from a displaced vertex. The ATLAS DV+$\slashed{E_T}$ search estimates the final events targeting at least one displaced vertex with jets and large missing transverse energy. So the signal processes which can give rise to the DV+$\slashed{E_T}$ signature at LHC in this framework are given as: 
\bea
p p &\to& \big ( \psi^{+} \psi^{-} \big)\to \overline{\xi}_2~\xi_2 +{\rm ~soft~pions} \to hh/hZ/ZZ + \overline{\chi}\chi \to \text{jets} + \overline{\chi}\chi ; \nonumber\\
  &\to& \big( \psi^{\pm}\xi_2 \big) \to \overline{\xi}_2~\xi_2 +{\rm ~soft~pion} \to hh/hZ/ZZ + \overline{\chi}\chi \to \text{jets} + \overline{\chi}\chi ; \nonumber \\
    &\to& \overline{\xi}_2~\xi_2 \to hh/hZ/ZZ + \overline{\chi}\chi \to \text{jets} + \overline{\chi}\chi.
\eea
Note that here we have only considered hadronic decay channels of $h$ and $Z$, since large hadronic branching yields a sizeable cross-section of the $pp\to \text{jets}+\overline{\chi}\chi$. The above signal cross-section mainly depends on the LLP (long-lived particle) mass, $m_{\xi_2} (\simeq m_{\Psi})$ and the Yukawa coupling, $Y$. The heavy neutral state, $\xi_2$ can decay inside ($ 1 \text{~mm}-100 ~\text{m}$) or outside the detector that crucially depends on the strength of $Y$.

An accurate prediction of discovery prospects of our proposed scenario requires proper recasting with the exact limits from CMS and ATLAS. One has to
perform a careful reconstruction and selection of events
employing suitable cuts, and by taking into account the
generator-level efficiency along with background estimation. The details of the recasting strategy of this singlet doublet model have been performed in ref.\cite{Calibbi:2018fqf}. In Figure  \ref{fig:llp} we extract the present and proposed exclusion bounds obtained on $m_{\xi_2}-Y$ plane from \cite{Calibbi:2018fqf}
and check the sensitivities of the benchmark points we used so far, implying different kinds of dark matter dynamics both in standard and non-standard cosmology.
The benchmark points FKD-I and FKD-II, which signify pure freeze-in in non-standard cosmology, are inside the 300 fb$^{-1}$ exclusion limit and can be traced in the future runs of ATLAS. One of two benchmark points (KD-II) is also likely to be probed in the next run of ATLAS. If tested, these would probably indicate the presence of modified cosmology with the values of non-standard cosmological parameters as listed in Table~\ref{tab:CIIa} and Table~\ref{tab:CIIb}. The benchmark points labelled as RD-I and RD-II are far outside the reach of ATLAS 300 fb$^{-1}$ run. For these reference points, the decay lengths of $\xi_2$ turns out to be very large with $c\tau_{\xi_2}\sim\mathcal{O}(1)$ km. Thus it is challenging to search $\xi_2$ with such long lifetimes at ATLAS and CMS with current and future sensitivities. The proposed MATHUSLA surface detector experiment~\cite{Curtin:2018mvb, MATHUSLA:2018bqv} could be capable of probing such a long-lived particle. We do not discuss this in detail and refer the readers to \cite{No:2019gvl}. 

\section{Dark matter self-interaction}\label{SI}
The present set-up resembles a two-component dark matter framework. We have seen in sections \ref{sec:CIIa} and \ref{sec:CIIB}, that $\chi$ could be the main component with almost 100$\%$ relic share owing to the presence of modified cosmology. Hence $\chi$, being adequately abundant in the present Universe, can be an ideal candidate for self-interacting dark matter. As earlier mentioned, few long-standing tensions between astrophysical observations and N-body stimulations for cold DM suggests the DM to be self-interacting with $1 ~\text{cm}^2/\text{gm}\lesssim \sigma/m_\chi\lesssim 10 ~\text{cm}^2/\text{gm}$ for DM relative velocity $(30 \lesssim v_d\lesssim 200)$ km/s \cite{Tulin:2017ara}. In the present framework, the self interaction can take place through $\phi$ mediation. We assume the $\chi$ to be symmetric in nature. The DM self-interaction processes are $\chi\chi\rightarrow\chi\chi$, $\overline{\chi}\overline{\chi}\rightarrow\overline{\chi}\overline{\chi}$ and $\overline{\chi}\chi\rightarrow\overline{\chi} \chi$. The nonrelativistic self-scattering in DM halos is conventionally described by a Yukawa potential,
\begin{align}
  V(r)=\pm \frac{\alpha_D^2}{4 \pi r}e^{-m_\phi r},
\end{align}
where ``$\pm$" stands for repulsive and attractive potentials. The parameter $\alpha_D$ is the analog of fine structure constant defined by $\alpha_D=\frac{\lambda^2}{4\pi}$. In general, a fermion dark matter can self-interacts via both scalar and vector portal. Note that the scalar
interactions are purely attractive, while a vector interaction could be both attractive or repulsive \cite{Tulin:2017ara}. In literature, uses of two kind of cross sections can be noticed namely (i) transfer cross section ($\sigma_T$) and (ii) viscosity cross section ($\sigma_V$), which are defined as follows~\cite{Tulin:2013teo, Tulin:2017ara, Buckley:2009in, Feng:2009hw, Mohapatra:2001sx, 1965itpg.book.....V}:
\begin{align}
  \sigma_T=\int d\Omega (1-\cos\theta)\frac{d\sigma}{d\Omega}~,~~~~\sigma_V=\int d\Omega\sin^2\theta\frac{d\sigma}{d\Omega}~~,
\end{align}
where $\theta$ is the scattering angle. The viscosity cross section has certain merits over the transfer one. For example, the $\sigma_V$ takes care of the divergences in both forward and backward scatterings in the DM halo. In addition, for self-interaction between identical particles, transfer cross sector fails. In view of this, we calculate the $\sigma_V$ in the present set up.

The description of the self-scattering could be of different natures (classical, semi-classical or quantum), parameterized by two dimensionless parameters $\kappa=\frac{m_\chi v_d}{m_\phi}$ and $\beta=\frac{2\alpha_\chi m_\phi}{m_\chi v_d^2}$ correlated to the momentum and strength of the potential relative to the kinetic energy. The system is known to be in the classical, semi-classical, and quantum regime for $\kappa\gg 1$,~$\kappa\gtrsim 1$ and $\kappa\lesssim 1$ respectively. The analytical form of $\sigma_V$ in classical and semi-classical regimes can be found in ref. \cite{Colquhoun:2020adl}.  

\begin{figure}[t]
\begin{center}
 \includegraphics[height=7.2cm,width=10.0cm]{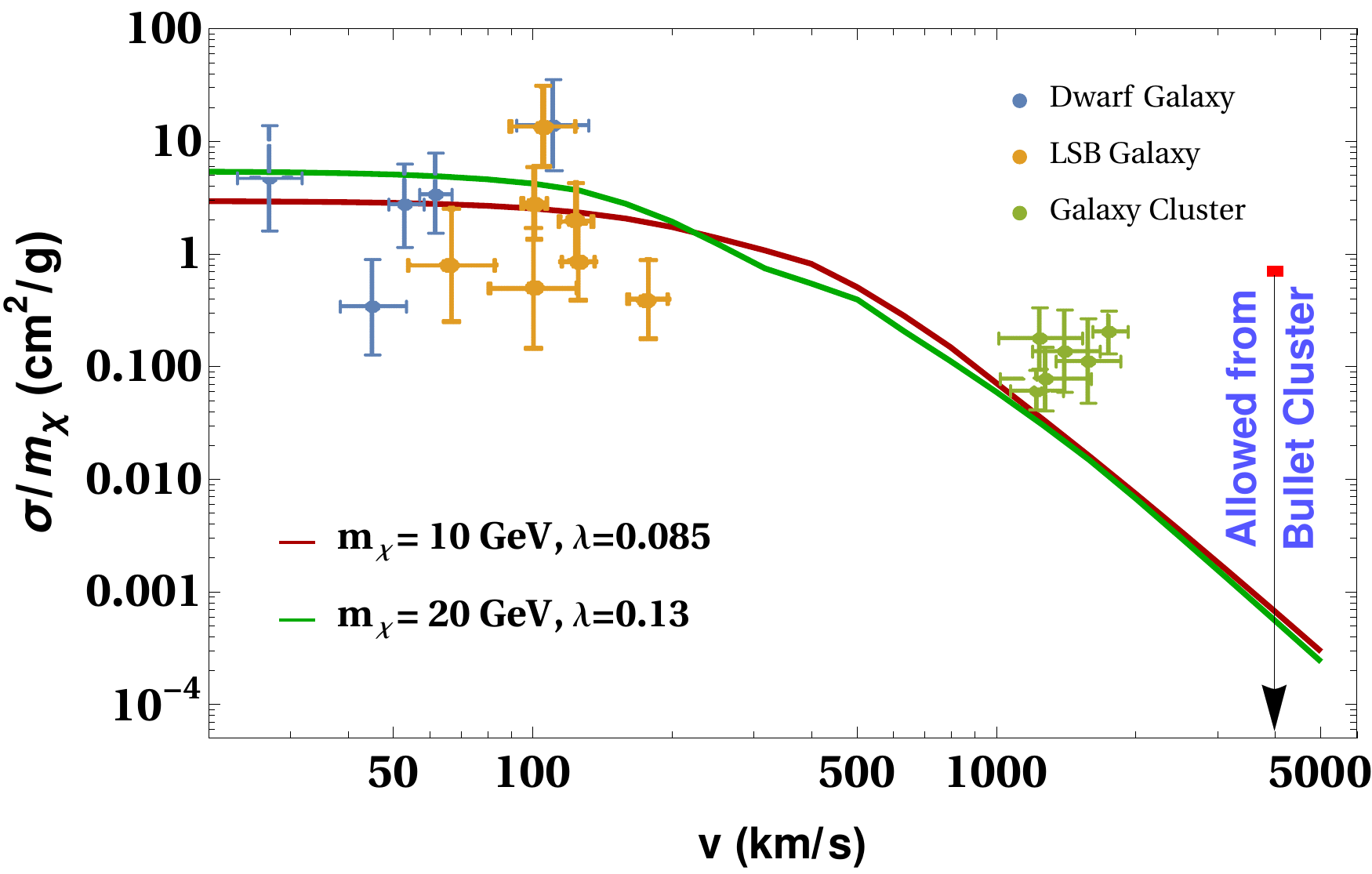}
 \end{center}
 \caption{Self-interaction cross section as function of velocity for the two different sets of $(m_\chi,\lambda)$ that have been used to describe the dark matter phenomenology in sections \ref{sec:CIIa} and \ref{sec:CIIB}. The black arrow indicates the constraint from the Bullet Cluster which is $\frac{\sigma}{m_\chi}< 0.7$ cm$^2$/g for $v=4000$ km/s~\cite{Randall:2008ppe, Tulin:2013teo}. We fix $m_\phi$ at 10 MeV. We also show the observational data from dwarfs (blue), low
surface brightness (LSB) galaxies (orange), and clusters
(green) as taken from \cite{Kaplinghat:2015aga}.}
 \label{fig:Siplot}
\end{figure}

For the quantum case (excluding the Born approximation for $2\beta\kappa^2\ll 1)$, in principle one needs to solve the Schrodinger equation by partial wave analysis in order to compute the differential cross section. The differential scattering cross section can be computed by,
\begin{align}
  \frac{d\sigma}{d\Omega}=\frac{1}{k^2}\left|(2l+1)e^{i\delta_l}P_l(\cos\theta)\sin\delta_l\right|^2.
\end{align}
where the phase shift for the partial wave is indicated by $\delta_l$. The Schrodinger equation for the radial wave function $R_l(r)$ of the reduced DM two-particle system is written as,
\begin{align}
  \frac{1}{r^2}\frac{\partial}{\partial r}\left(r^2 \frac{d R_l}{d r}\right)+\left(k^2-\frac{l(l+1)}{r^2}-2\mu V(r)\right)R_l=0,
\end{align}
where $\mu=m_\chi/2$ is the reduced mass and $k=\mu v_d$. 
Then by obtaining $\delta_l$ from the asymptotic form of the radial function $R_l(r)$, one can reach at the approximated expression for the viscosity cross section:
\begin{align}
  \sigma_V m_\phi^2\simeq \frac{4\pi}{\kappa^2}\int_0^\infty dl\frac{(l+\frac{1}{2})(l+\frac{3}{2})}{2l+2} ~\sin^22\delta^\prime\left(l+\frac{1}{2}\right),
\end{align}
where $\delta^\prime\left(l+1/2\right)\simeq \delta_{l+1}-\delta_l$. 
The analytical computation of $\sigma_V$ looks unlikely in the above case, where both quantum and non-resonant effects are important. However, for $\kappa\ll 1$, the S-wave scattering dominates, and the Hulthen potential can be implemented to find the $\sigma_V$ analytically as indicated in \cite{Colquhoun:2020adl}. 

In Figure  \ref{fig:Siplot}, we estimate the velocity dependence of the viscosity cross section $\sigma_V$ considering two sets of $(m_\chi,\lambda)$ as used for the DM analysis earlier in section \ref{sec:CIIa} and section \ref{sec:CIIB} with $m_\phi$= 10 MeV. Note that for both the benchmark points, we could obtain nearly same percent relic abundance for the dark matter candidate $\chi$ in considering non-standard cosmology namely kination domination or faster than kination domination. For both the $(m_\chi,\lambda)$ sets, the parameter $\frac{\sigma_v}{m_\chi}$ remains in the range (1-10) cm$^2$/g at $v=(30-200)$ km/s.
We also found that these points also satisfy the bound $\frac{\sigma_v}{m_\chi}< 0.7$ cm$^2$/g at $v=4000$ km/s, arising from the observations in Bullet Cluster galaxies.
In the case of RD dominated Universe, for the two benchmark points under our discussion, $\chi$ is not adequately abundant in the present scenario, and hence the solution of small structure anomalies remains unlikely. 
However, the assumption of a modified cosmology before BBN turns instrumental in the simultaneous realisation of adequately abundant $\chi$ component dark matter and sufficient self-interaction to alleviate the small scale anomalies of the Universe.

\section{Summary and Conclusion}\label{conclusion}
The singlet doublet model is a simple particle extension of SM, providing a viable dark matter candidate from mixing of doublet and singlet fermions after the electroweak symmetry breaking. Different variants of this scenario are extensively studied for its enriched dark sector and collider signatures. In this work, we have examined whether a non-thermally produced and adequately abundant GeV scale fermion doublet dark matter candidate has the potential to be probed at colliders. We also discuss the prospect of alleviating the small scale structure anomalies of the Universe. 

We have minimally extended the singlet doublet dark matter model with a MeV scale singlet scalar, which mediates the dark matter self-interaction. The singlet scalar is stable in Universe lifetime since it has no decay mode. The fermion dark matter has a non-thermal origin and can be produced adequately in the early phase of the Universe. A strong self-interaction, in general, prefers a sizable non-gravitational interaction strength between the dark matter and the mediator particle. With such a notion, our computation indicates the formation of internal dark thermal equilibrium when the conversion process from the fermion dark matter to scalar mediator turns so efficient that it significantly suppresses relic abundance for our dark matter candidate. In fact, the mediator particle emerges to be way more abundant in the present Universe.
The process as mentioned above is prominent in a standard radiation-dominated Universe, and we remain unsuccessful in obtaining the singlet doublet fermion dark matter as the main component.
 
We adopt two specific non-standard cosmological scenarios such as kination and faster than kination-dominated early Universe to circumvent this issue. The motivation for such choices is to suppress the conversion process inside the dark sector. We have found that proper tunings of non-standard cosmological parameters completely alter the evolution patterns of the number densities for the dark sector particles. In fact, the benchmark points which had earlier manifested a reannihilating pattern in the RD dominated Universe depict a pure freeze-in scenario in the non-standard Universe with negligible mediator abundance in the present Universe. In addition, for some choices of the non-standard cosmological parameters, the dark sector still goes through internal thermal equilibrium. However, that does not cause colossal depletion to the dark matter relic, unlike the case considering radiation domination. 

In short, the presence of a modified cosmology helps realise the fermion-dark matter as the main component that is adequately abundant in the present Universe.
An exciting consequence comes in terms of collider constraints, where we found that the displaced Vertex signature can provide the robust exclusion bound on the GeV scale DM parameter space. Some of our benchmark points are already within the projected exclusion limit that can be tested by LHC in the next run. We have further demonstrated that the realised parameter space can indeed generate velocity-dependent sufficient self-interactions (consistent with the bounds from observations at bullet cluster galaxies) with a MeV scale mediator. 

We conclude with the comment that, in general, a freeze-in dark matter being feebly coupled to the visible sector is extremely hard to track at experiments. Interestingly, the proposed framework of singlet doublet freeze-in GeV scale DM has ample scopes to be indirectly probed {\it e.g.} in the astrophysical experiments at galaxy scales due to the strong self-interaction of the fermion dark matter as well as in the collider experiments by virtue of modified cosmological theory.  


\appendix

 \section{Cross section}
 The analytical expression for the cross section of $\chi\chi\rightarrow\phi\phi$ process is given by,
 \beqn
 \sigma_{\chi\chi\rightarrow \phi\phi}&=&\frac{1}{32\pi s(s-4m_{\chi}^2)}\times\Bigg\{\frac{\ln B}{s-2m_{\phi}^2} \Big[
 \lambda^4s^2 -4\lambda^4\left(m_{\phi}^2-4m_{\chi}^2 \right)s\nn +6\lambda^4m_{\phi}^4 - 16\lambda^4m_{\phi}^2m_{\chi}^2
 \eeqn

 \beqn
 &~&-32\lambda^4m_{\chi}^2
 \Big]
 -\frac{\sqrt{(s-4m_{\chi}^2)(s-4m_{\phi}^2)}}{m_{\chi}^2s-4m_{\phi}^2m_{\chi}^2+m_{\phi}^4}\Big[\lambda^4\left(2m_{\chi}^2(s+8m_{\chi}^2) -16m_{\phi}^2m_{\chi}^2 +3m_{\phi}^4\right)
 \Big]
 \Bigg\}\,,\label{eq:xsection_chichiphiphi}\nn\\
 \eeqn
 where 
 \beq
 B~\equiv~ \frac{s-2m_{\phi}^2+\sqrt{(s-4m_{\chi}^2)(s-4m_{\phi}^2)}}{s-2m_{\phi}^2-\sqrt{(s-4m_{\chi})(s-4m_{\phi}^2)}} \,.
 \eeq

 \section{Decay widths}
 Before EW symmetry breaking the production of dark matter occurs from $\Psi\rightarrow H\chi$ decay while in the post EW symmetry breaking era, decays such as $\xi_2\rightarrow h \chi$, $\xi_2\rightarrow Z \chi$ and $\psi^\pm\rightarrow W^\pm \chi$ turn active. The analytical expression for the decay widths are read as,
 \begin{align}
  &\Gamma_{\xi_2\rightarrow h \chi}=\frac{Y^2}{32\pi m_{\xi_2}^3}\left\{(m_{\xi_2}+m_\chi)^2-m_h^2\right\}F(m_{\xi_2},m_\chi,m_h),\\
  &\Gamma(\xi_2\rightarrow Z\chi)=\frac{Y^2}{32\pi}\frac{\left\{(m_{\xi_2}-m_\chi)^2-m_Z^2\right\}\left\{(m_{\xi_2}-m_\chi)^2+2 m_Z^2\right\}}{m_{\xi_2}^3(m_{\xi_2}-m_{\chi})^2}F(m_{\xi_2},m_\chi,m_Z),\\
  &\Gamma(\psi^\pm\rightarrow W^\pm\chi)=\frac{Y^2}{32\pi}\frac{\{(m_\Psi-m_\chi)^2-m_W^2\}\{(m_\Psi+m_\chi)^2+2 m_W^2\}}{m_\Psi^3(m_\Psi-m_\chi)^2}F(m_\Psi,m_\chi,m_W),
 \end{align}
 where $F(x,y,z)=\sqrt{x^4+y^4+z^4-2x^2y^2-2x^2z^2-2y^2z^2}$. We also note that
 \begin{align}
 \Gamma_{\Psi\rightarrow H\chi}\simeq 4 ~ \Gamma (\xi_2\rightarrow h \chi).
 \end{align}
 
 \section{Energy Transfer rate from SM to dark sector}\label{sec:energy_transfer}
 The aim of this appendix is to determine an analytical form of the energy transfer rate for two body decay of $\Psi_i$ appearing in R.H.S of Eq.(\ref{eq:darkTempCI}).
 We start with the Boltzmann equation for the phase-space distribution of a particle species $\Psi_i$:
 \beq
 \frac{\partial f_{\Psi_i}(p_{\Psi_i},t)}{\partial t}~=~H(t)p_{\Psi_i}\frac{\partial f_{\Psi_i}(p_{\Psi_i},t)}{\partial p_{\Psi_i}} + C[f]\,,\label{eq:boltzmann_f},
 \eeq
 where $C[f]$ is the collision operator for tree level $\Psi_i$ decay.
 For a specific process $\Psi_i\rightarrow a+b$ process, the collision term takes the following form,
 \beqn
 C[f]&=&-\frac{1}{2E_{\Psi_i}}\int d\Pi_a d\Pi_b (2\pi)^4 \delta^{(4)}(p_{\Psi_i}-p_a-p_b)\times\bigg[\abs{\mathcal{M}_{\Psi_i\rightarrow a+b}}^2f_{\Psi_i}(1\pm f_a)(1\pm f_b)\nn\\
 &~&~~~~ - \abs{\mathcal{M}_{a+b\rightarrow  \Psi}}^2f_a f_b (1\pm f_{\Psi_i}) \bigg]\,,\label{eq:collision}
 \eeqn
 where $d\Pi_i=\frac{d^3p_{i}}{(2\pi)^3 2 E_i}$, and ``$\pm$'' symbolises the Bose-enhancement and Pauli-blocking factors. We have used the Maxwell-Boltzmann distribution approximation and  use $f_i\approx e^{-E_i/T_i}$ and $1\pm f_i\approx 1$  to simplify the calculation.
 The energy-transfer rate from the SM sector to the dark sector is obtained by 
 \beq
 \frac{d \rho_{\Psi_i}}{d t}~=~-3H(P_{\Psi_i}+\rho_{\Psi_i})+\int \frac{d^3p_{\Psi_i}}{(2\pi)^3} E_{\Psi_i} C[f]\,.
 \eeq
 In case for a decay, we write
 \begin{align}
  \int \frac{d^3p_{\psi}}{(2\pi)^3} E_{\psi} C[f]=n_{\Psi_i} \mathcal{P}_{\Psi_i},
 \end{align}
where we obtain $n_{\Psi_i} \mathcal{P}_{\Psi_i}=n_{\Psi_i} m_{\Psi_i}\Gamma_{\Psi_i}$.

\acknowledgments

PK and SS gratefully acknowledge the HPC resources (Vikram-100 HPC) and TDP project at PRL for the computations. AKS appreciates Sougata Ganguly
for several discussions during the course of work. AKS is
supported by NPDF grant PDF/2020/000797 from Sci-
ence and Engineering Research Board, Government of
India. PG would like to acknowledge the support from DAE, India for the Regional Centre for Accelerator based Particle Physics (RECAPP), Harish Chandra Research Institute.

\paragraph{Note added:} Shortly before submitting this manuscript, the preprint \cite{Borah:2021rbx} appeared on the arXiv, which also considers the self-interacting dark matter in an enlarged variant of the singlet doublet framework. The dark sector is purely thermal in their case under the assumption of a standard radiation dominated Universe.
In contrast, we have explored a freeze-in production of a hidden dark sector considering a few different variants of cosmological scenarios.

\bibliographystyle{JHEP}
\bibliography{ref}  

\providecommand{\href}[2]{#2}\begingroup\raggedright\begin{thebibliography}{10}

\bibitem{Hall:2009bx}
L.~J. Hall, K.~Jedamzik, J.~March-Russell and S.~M. West, \emph{{Freeze-In
  Production of FIMP Dark Matter}},
  \href{https://doi.org/10.1007/JHEP03(2010)080}{\emph{JHEP} {\bfseries 03}
  (2010) 080} [\href{https://arxiv.org/abs/0911.1120}{{\ttfamily 0911.1120}}].

\bibitem{1991ApJ...378..496D}
J.~{Dubinski} and R.~G. {Carlberg}, \emph{{The Structure of Cold Dark Matter
  Halos}}, \href{https://doi.org/10.1086/170451}{\emph{Astrophysical Journal}
  {\bfseries 378} (1991) 496}.

\bibitem{Flores:1994gz}
R.~A. Flores and J.~R. Primack, \emph{{Observational and theoretical
  constraints on singular dark matter halos}},
  \href{https://doi.org/10.1086/187350}{\emph{Astrophys. J. Lett.} {\bfseries
  427} (1994) L1} [\href{https://arxiv.org/abs/astro-ph/9402004}{{\ttfamily
  astro-ph/9402004}}].

\bibitem{Tulin:2013teo}
S.~Tulin, H.-B. Yu and K.~M. Zurek, \emph{{Beyond Collisionless Dark Matter:
  Particle Physics Dynamics for Dark Matter Halo Structure}},
  \href{https://doi.org/10.1103/PhysRevD.87.115007}{\emph{Phys. Rev. D}
  {\bfseries 87} (2013) 115007}
  [\href{https://arxiv.org/abs/1302.3898}{{\ttfamily 1302.3898}}].

\bibitem{2011}
M.~Boylan-Kolchin, J.~S. Bullock and M.~Kaplinghat, \emph{Too big to fail? the
  puzzling darkness of massive milky way subhaloes},
  \href{https://doi.org/10.1111/j.1745-3933.2011.01074.x}{\emph{Monthly Notices
  of the Royal Astronomical Society: Letters} {\bfseries 415} (2011)
  L40–L44}.

\bibitem{2010}
M.~A. Zwaan, M.~J. Meyer and L.~Staveley-Smith, \emph{The velocity function of
  gas-rich galaxies},
  \href{https://doi.org/10.1111/j.1365-2966.2009.16188.x}{\emph{Monthly Notices
  of the Royal Astronomical Society} {\bfseries 403} (2010) 1969–1977}.

\bibitem{Spergel:1999mh}
D.~N. Spergel and P.~J. Steinhardt, \emph{{Observational evidence for
  selfinteracting cold dark matter}},
  \href{https://doi.org/10.1103/PhysRevLett.84.3760}{\emph{Phys. Rev. Lett.}
  {\bfseries 84} (2000) 3760}
  [\href{https://arxiv.org/abs/astro-ph/9909386}{{\ttfamily
  astro-ph/9909386}}].

\bibitem{Bullock:2017xww}
J.~S. Bullock and M.~Boylan-Kolchin, \emph{{Small-Scale Challenges to the
  $\Lambda$CDM Paradigm}},
  \href{https://doi.org/10.1146/annurev-astro-091916-055313}{\emph{Ann. Rev.
  Astron. Astrophys.} {\bfseries 55} (2017) 343}
  [\href{https://arxiv.org/abs/1707.04256}{{\ttfamily 1707.04256}}].

\bibitem{Klypin:1999uc}
A.~A. Klypin, A.~V. Kravtsov, O.~Valenzuela and F.~Prada, \emph{{Where are the
  missing Galactic satellites?}},
  \href{https://doi.org/10.1086/307643}{\emph{Astrophys. J.} {\bfseries 522}
  (1999) 82} [\href{https://arxiv.org/abs/astro-ph/9901240}{{\ttfamily
  astro-ph/9901240}}].

\bibitem{Moore:1999nt}
B.~Moore, S.~Ghigna, F.~Governato, G.~Lake, T.~R. Quinn, J.~Stadel et~al.,
  \emph{{Dark matter substructure within galactic halos}},
  \href{https://doi.org/10.1086/312287}{\emph{Astrophys. J. Lett.} {\bfseries
  524} (1999) L19} [\href{https://arxiv.org/abs/astro-ph/9907411}{{\ttfamily
  astro-ph/9907411}}].

\bibitem{Moore:1999gc}
B.~Moore, T.~R. Quinn, F.~Governato, J.~Stadel and G.~Lake, \emph{{Cold
  collapse and the core catastrophe}},
  \href{https://doi.org/10.1046/j.1365-8711.1999.03039.x}{\emph{Mon. Not. Roy.
  Astron. Soc.} {\bfseries 310} (1999) 1147}
  [\href{https://arxiv.org/abs/astro-ph/9903164}{{\ttfamily
  astro-ph/9903164}}].

\bibitem{Springel:2008cc}
V.~Springel, J.~Wang, M.~Vogelsberger, A.~Ludlow, A.~Jenkins, A.~Helmi et~al.,
  \emph{{The Aquarius Project: the subhalos of galactic halos}},
  \href{https://doi.org/10.1111/j.1365-2966.2008.14066.x}{\emph{Mon. Not. Roy.
  Astron. Soc.} {\bfseries 391} (2008) 1685}
  [\href{https://arxiv.org/abs/0809.0898}{{\ttfamily 0809.0898}}].

\bibitem{Randall:2008ppe}
S.~W. Randall, M.~Markevitch, D.~Clowe, A.~H. Gonzalez and M.~Bradac,
  \emph{{Constraints on the Self-Interaction Cross-Section of Dark Matter from
  Numerical Simulations of the Merging Galaxy Cluster 1E 0657-56}},
  \href{https://doi.org/10.1086/587859}{\emph{Astrophys. J.} {\bfseries 679}
  (2008) 1173} [\href{https://arxiv.org/abs/0704.0261}{{\ttfamily 0704.0261}}].

\bibitem{Kouvaris:2014uoa}
C.~Kouvaris, I.~M. Shoemaker and K.~Tuominen, \emph{{Self-Interacting Dark
  Matter through the Higgs Portal}},
  \href{https://doi.org/10.1103/PhysRevD.91.043519}{\emph{Phys. Rev. D}
  {\bfseries 91} (2015) 043519}
  [\href{https://arxiv.org/abs/1411.3730}{{\ttfamily 1411.3730}}].

\bibitem{Kainulainen:2015sva}
K.~Kainulainen, K.~Tuominen and V.~Vaskonen, \emph{{Self-interacting dark
  matter and cosmology of a light scalar mediator}},
  \href{https://doi.org/10.1103/PhysRevD.93.015016}{\emph{Phys. Rev. D}
  {\bfseries 93} (2016) 015016}
  [\href{https://arxiv.org/abs/1507.04931}{{\ttfamily 1507.04931}}].

\bibitem{Borah:2021jzu}
D.~Borah, M.~Dutta, S.~Mahapatra and N.~Sahu, \emph{{Muon
  (g \ensuremath{-} 2) and XENON1T excess with boosted dark matter in
  L\ensuremath{\mu} \ensuremath{-} L\ensuremath{\tau} model}},
  \href{https://doi.org/10.1016/j.physletb.2021.136577}{\emph{Phys. Lett. B}
  {\bfseries 820} (2021) 136577}
  [\href{https://arxiv.org/abs/2104.05656}{{\ttfamily 2104.05656}}].

\bibitem{Dutta:2021wbn}
M.~Dutta, S.~Mahapatra, D.~Borah and N.~Sahu, \emph{{Self-interacting Inelastic
  Dark Matter in the light of XENON1T excess}},
  \href{https://doi.org/10.1103/PhysRevD.103.095018}{\emph{Phys. Rev. D}
  {\bfseries 103} (2021) 095018}
  [\href{https://arxiv.org/abs/2101.06472}{{\ttfamily 2101.06472}}].

\bibitem{Borah:2021yek}
D.~Borah, M.~Dutta, S.~Mahapatra and N.~Sahu, \emph{{Boosted Self-Interacting
  Dark Matter and XENON1T Excess}},
  \href{https://arxiv.org/abs/2107.13176}{{\ttfamily 2107.13176}}.

\bibitem{Duch:2017khv}
M.~Duch, B.~Grzadkowski and D.~Huang, \emph{{Strongly self-interacting vector
  dark matter via freeze-in}},
  \href{https://doi.org/10.1007/JHEP01(2018)020}{\emph{JHEP} {\bfseries 01}
  (2018) 020} [\href{https://arxiv.org/abs/1710.00320}{{\ttfamily
  1710.00320}}].

\bibitem{Du:2020avz}
Y.~Du, F.~Huang, H.-L. Li and J.-H. Yu, \emph{{Freeze-in Dark Matter from
  Secret Neutrino Interactions}},
  \href{https://doi.org/10.1007/JHEP12(2020)207}{\emph{JHEP} {\bfseries 12}
  (2020) 207} [\href{https://arxiv.org/abs/2005.01717}{{\ttfamily
  2005.01717}}].

\bibitem{DEramo:2007anh}
F.~D'Eramo, \emph{{Dark matter and Higgs boson physics}},
  \href{https://doi.org/10.1103/PhysRevD.76.083522}{\emph{Phys. Rev. D}
  {\bfseries 76} (2007) 083522}
  [\href{https://arxiv.org/abs/0705.4493}{{\ttfamily 0705.4493}}].

\bibitem{Enberg:2007rp}
R.~Enberg, P.~J. Fox, L.~J. Hall, A.~Y. Papaioannou and M.~Papucci, \emph{{LHC
  and dark matter signals of improved naturalness}},
  \href{https://doi.org/10.1088/1126-6708/2007/11/014}{\emph{JHEP} {\bfseries
  11} (2007) 014} [\href{https://arxiv.org/abs/0706.0918}{{\ttfamily
  0706.0918}}].

\bibitem{Cohen:2011ec}
T.~Cohen, J.~Kearney, A.~Pierce and D.~Tucker-Smith, \emph{{Singlet-Doublet
  Dark Matter}}, \href{https://doi.org/10.1103/PhysRevD.85.075003}{\emph{Phys.
  Rev. D} {\bfseries 85} (2012) 075003}
  [\href{https://arxiv.org/abs/1109.2604}{{\ttfamily 1109.2604}}].

\bibitem{Cheung:2013dua}
C.~Cheung and D.~Sanford, \emph{{Simplified Models of Mixed Dark Matter}},
  \href{https://doi.org/10.1088/1475-7516/2014/02/011}{\emph{JCAP} {\bfseries
  02} (2014) 011} [\href{https://arxiv.org/abs/1311.5896}{{\ttfamily
  1311.5896}}].

\bibitem{Restrepo:2015ura}
D.~Restrepo, A.~Rivera, M.~S\'anchez-Pel\'aez, O.~Zapata and W.~Tangarife,
  \emph{{Radiative Neutrino Masses in the Singlet-Doublet Fermion Dark Matter
  Model with Scalar Singlets}},
  \href{https://doi.org/10.1103/PhysRevD.92.013005}{\emph{Phys. Rev. D}
  {\bfseries 92} (2015) 013005}
  [\href{https://arxiv.org/abs/1504.07892}{{\ttfamily 1504.07892}}].

\bibitem{Calibbi:2015nha}
L.~Calibbi, A.~Mariotti and P.~Tziveloglou, \emph{{Singlet-Doublet Model: Dark
  matter searches and LHC constraints}},
  \href{https://doi.org/10.1007/JHEP10(2015)116}{\emph{JHEP} {\bfseries 10}
  (2015) 116} [\href{https://arxiv.org/abs/1505.03867}{{\ttfamily
  1505.03867}}].

\bibitem{Bhattacharya:2015qpa}
S.~Bhattacharya, N.~Sahoo and N.~Sahu, \emph{{Minimal vectorlike leptonic dark
  matter and signatures at the LHC}},
  \href{https://doi.org/10.1103/PhysRevD.93.115040}{\emph{Phys. Rev. D}
  {\bfseries 93} (2016) 115040}
  [\href{https://arxiv.org/abs/1510.02760}{{\ttfamily 1510.02760}}].

\bibitem{Yaguna:2015mva}
C.~E. Yaguna, \emph{{Singlet-Doublet Dirac Dark Matter}},
  \href{https://doi.org/10.1103/PhysRevD.92.115002}{\emph{Phys. Rev. D}
  {\bfseries 92} (2015) 115002}
  [\href{https://arxiv.org/abs/1510.06151}{{\ttfamily 1510.06151}}].

\bibitem{Horiuchi:2016tqw}
S.~Horiuchi, O.~Macias, D.~Restrepo, A.~Rivera, O.~Zapata and H.~Silverwood,
  \emph{{The Fermi-LAT gamma-ray excess at the Galactic Center in the
  singlet-doublet fermion dark matter model}},
  \href{https://doi.org/10.1088/1475-7516/2016/03/048}{\emph{JCAP} {\bfseries
  03} (2016) 048} [\href{https://arxiv.org/abs/1602.04788}{{\ttfamily
  1602.04788}}].

\bibitem{Banerjee:2016hsk}
S.~Banerjee, S.~Matsumoto, K.~Mukaida and Y.-L.~S. Tsai, \emph{{WIMP Dark
  Matter in a Well-Tempered Regime: A case study on Singlet-Doublets Fermionic
  WIMP}}, \href{https://doi.org/10.1007/JHEP11(2016)070}{\emph{JHEP} {\bfseries
  11} (2016) 070} [\href{https://arxiv.org/abs/1603.07387}{{\ttfamily
  1603.07387}}].

\bibitem{Abe:2017glm}
T.~Abe, \emph{{Effect of CP violation in the singlet-doublet dark matter
  model}}, \href{https://doi.org/10.1016/j.physletb.2017.05.048}{\emph{Phys.
  Lett. B} {\bfseries 771} (2017) 125}
  [\href{https://arxiv.org/abs/1702.07236}{{\ttfamily 1702.07236}}].

\bibitem{Maru:2017pwl}
N.~Maru, N.~Okada and S.~Okada, \emph{{Fermionic Minimal Dark Matter in 5D
  Gauge-Higgs Unification}},
  \href{https://doi.org/10.1103/PhysRevD.96.115023}{\emph{Phys. Rev. D}
  {\bfseries 96} (2017) 115023}
  [\href{https://arxiv.org/abs/1801.00686}{{\ttfamily 1801.00686}}].

\bibitem{Maru:2017otg}
N.~Maru, T.~Miyaji, N.~Okada and S.~Okada, \emph{{Fermion Dark Matter in
  Gauge-Higgs Unification}},
  \href{https://doi.org/10.1007/JHEP07(2017)048}{\emph{JHEP} {\bfseries 07}
  (2017) 048} [\href{https://arxiv.org/abs/1704.04621}{{\ttfamily
  1704.04621}}].

\bibitem{Bhattacharya:2017sml}
S.~Bhattacharya, N.~Sahoo and N.~Sahu, \emph{{Singlet-Doublet Fermionic Dark
  Matter, Neutrino Mass and Collider Signatures}},
  \href{https://doi.org/10.1103/PhysRevD.96.035010}{\emph{Phys. Rev. D}
  {\bfseries 96} (2017) 035010}
  [\href{https://arxiv.org/abs/1704.03417}{{\ttfamily 1704.03417}}].

\bibitem{Xiang:2017yfs}
Q.-F. Xiang, X.-J. Bi, P.-F. Yin and Z.-H. Yu, \emph{{Exploring Fermionic Dark
  Matter via Higgs Boson Precision Measurements at the Circular Electron
  Positron Collider}},
  \href{https://doi.org/10.1103/PhysRevD.97.055004}{\emph{Phys. Rev. D}
  {\bfseries 97} (2018) 055004}
  [\href{https://arxiv.org/abs/1707.03094}{{\ttfamily 1707.03094}}].

\bibitem{Esch:2018ccs}
S.~Esch, M.~Klasen and C.~E. Yaguna, \emph{{A singlet doublet dark matter model
  with radiative neutrino masses}},
  \href{https://doi.org/10.1007/JHEP10(2018)055}{\emph{JHEP} {\bfseries 10}
  (2018) 055} [\href{https://arxiv.org/abs/1804.03384}{{\ttfamily
  1804.03384}}].

\bibitem{Arcadi:2018pfo}
G.~Arcadi, \emph{{2HDM portal for Singlet-Doublet Dark Matter}},
  \href{https://doi.org/10.1140/epjc/s10052-018-6327-6}{\emph{Eur. Phys. J. C}
  {\bfseries 78} (2018) 864}
  [\href{https://arxiv.org/abs/1804.04930}{{\ttfamily 1804.04930}}].

\bibitem{DuttaBanik:2018emv}
A.~Dutta~Banik, A.~K. Saha and A.~Sil, \emph{{Scalar assisted singlet doublet
  fermion dark matter model and electroweak vacuum stability}},
  \href{https://doi.org/10.1103/PhysRevD.98.075013}{\emph{Phys. Rev. D}
  {\bfseries 98} (2018) 075013}
  [\href{https://arxiv.org/abs/1806.08080}{{\ttfamily 1806.08080}}].

\bibitem{Bhattacharya:2018fus}
S.~Bhattacharya, P.~Ghosh, N.~Sahoo and N.~Sahu, \emph{{Mini Review on
  Vector-Like Leptonic Dark Matter, Neutrino Mass, and Collider Signatures}},
  \href{https://doi.org/10.3389/fphy.2019.00080}{\emph{Front. in Phys.}
  {\bfseries 7} (2019) 80} [\href{https://arxiv.org/abs/1812.06505}{{\ttfamily
  1812.06505}}].

\bibitem{Fiaschi:2018rky}
J.~Fiaschi, M.~Klasen and S.~May, \emph{{Singlet-doublet fermion and triplet
  scalar dark matter with radiative neutrino masses}},
  \href{https://doi.org/10.1007/JHEP05(2019)015}{\emph{JHEP} {\bfseries 05}
  (2019) 015} [\href{https://arxiv.org/abs/1812.11133}{{\ttfamily
  1812.11133}}].

\bibitem{Barman:2019tuo}
B.~Barman, S.~Bhattacharya, P.~Ghosh, S.~Kadam and N.~Sahu, \emph{{Fermion Dark
  Matter with Scalar Triplet at Direct and Collider Searches}},
  \href{https://doi.org/10.1103/PhysRevD.100.015027}{\emph{Phys. Rev. D}
  {\bfseries 100} (2019) 015027}
  [\href{https://arxiv.org/abs/1902.01217}{{\ttfamily 1902.01217}}].

\bibitem{Restrepo:2019soi}
D.~Restrepo, A.~Rivera and W.~Tangarife, \emph{{Singlet-Doublet Dirac Dark
  Matter and Neutrino Masses}},
  \href{https://doi.org/10.1103/PhysRevD.100.035029}{\emph{Phys. Rev. D}
  {\bfseries 100} (2019) 035029}
  [\href{https://arxiv.org/abs/1906.09685}{{\ttfamily 1906.09685}}].

\bibitem{Barman:2019aku}
B.~Barman, D.~Borah, P.~Ghosh and A.~K. Saha, \emph{{Flavoured gauge extension
  of singlet-doublet fermionic dark matter: neutrino mass, high scale validity
  and collider signatures}},
  \href{https://doi.org/10.1007/JHEP10(2019)275}{\emph{JHEP} {\bfseries 10}
  (2019) 275} [\href{https://arxiv.org/abs/1907.10071}{{\ttfamily
  1907.10071}}].

\bibitem{Konar:2020wvl}
P.~Konar, A.~Mukherjee, A.~K. Saha and S.~Show, \emph{{Linking pseudo-Dirac
  dark matter to radiative neutrino masses in a singlet-doublet scenario}},
  \href{https://doi.org/10.1103/PhysRevD.102.015024}{\emph{Phys. Rev. D}
  {\bfseries 102} (2020) 015024}
  [\href{https://arxiv.org/abs/2001.11325}{{\ttfamily 2001.11325}}].

\bibitem{Konar:2020vuu}
P.~Konar, A.~Mukherjee, A.~K. Saha and S.~Show, \emph{{A dark clue to seesaw
  and leptogenesis in a pseudo-Dirac singlet doublet scenario with
  (non)standard cosmology}},
  \href{https://doi.org/10.1007/JHEP03(2021)044}{\emph{JHEP} {\bfseries 03}
  (2021) 044} [\href{https://arxiv.org/abs/2007.15608}{{\ttfamily
  2007.15608}}].

\bibitem{Dutta:2020xwn}
M.~Dutta, S.~Bhattacharya, P.~Ghosh and N.~Sahu, \emph{{Singlet-Doublet
  Majorana Dark Matter and Neutrino Mass in a minimal Type-I Seesaw Scenario}},
  \href{https://doi.org/10.1088/1475-7516/2021/03/008}{\emph{JCAP} {\bfseries
  03} (2021) 008} [\href{https://arxiv.org/abs/2009.00885}{{\ttfamily
  2009.00885}}].

\bibitem{Barman:2021ifu}
B.~Barman, P.~Ghosh, F.~S. Queiroz and A.~K. Saha, \emph{{Scalar multiplet dark
  matter in a fast expanding Universe: Resurrection of the desert region}},
  \href{https://doi.org/10.1103/PhysRevD.104.015040}{\emph{Phys. Rev. D}
  {\bfseries 104} (2021) 015040}
  [\href{https://arxiv.org/abs/2101.10175}{{\ttfamily 2101.10175}}].

\bibitem{Borah:2021khc}
D.~Borah, M.~Dutta, S.~Mahapatra and N.~Sahu, \emph{{Lepton Anomalous Magnetic
  Moment with Singlet-Doublet Fermion Dark Matter in Scotogenic
  $U(1)_{L_{\mu}-L_{\tau}}$ Model}},
  \href{https://arxiv.org/abs/2109.02699}{{\ttfamily 2109.02699}}.

\bibitem{Calibbi:2018fqf}
L.~Calibbi, L.~Lopez-Honorez, S.~Lowette and A.~Mariotti,
  \emph{{Singlet-Doublet Dark Matter Freeze-in: LHC displaced signatures versus
  cosmology}}, \href{https://doi.org/10.1007/JHEP09(2018)037}{\emph{JHEP}
  {\bfseries 09} (2018) 037}
  [\href{https://arxiv.org/abs/1805.04423}{{\ttfamily 1805.04423}}].

\bibitem{No:2019gvl}
J.~M. No, P.~Tunney and B.~Zaldivar, \emph{{Probing Dark Matter freeze-in with
  long-lived particle signatures: MATHUSLA, HL-LHC and FCC-hh}},
  \href{https://doi.org/10.1007/JHEP03(2020)022}{\emph{JHEP} {\bfseries 03}
  (2020) 022} [\href{https://arxiv.org/abs/1908.11387}{{\ttfamily
  1908.11387}}].

\bibitem{Zhu:2021pad}
B.~Zhu and M.~Abdughani, \emph{{Thermal Relic of Self-Interacting Dark Matter
  with Retarded Decay of Mediator}},
  \href{https://arxiv.org/abs/2103.06050}{{\ttfamily 2103.06050}}.

\bibitem{Krnjaic:2017tio}
G.~Krnjaic, \emph{{Freezing In, Heating Up, and Freezing Out: Predictive
  Nonthermal Dark Matter and Low-Mass Direct Detection}},
  \href{https://doi.org/10.1007/JHEP10(2018)136}{\emph{JHEP} {\bfseries 10}
  (2018) 136} [\href{https://arxiv.org/abs/1711.11038}{{\ttfamily
  1711.11038}}].

\bibitem{Berger:2018xyd}
J.~Berger, D.~Croon, S.~El~Hedri, K.~Jedamzik, A.~Perko and D.~G.~E. Walker,
  \emph{{Dark matter amnesia in out-of-equilibrium scenarios}},
  \href{https://doi.org/10.1088/1475-7516/2019/02/051}{\emph{JCAP} {\bfseries
  02} (2019) 051} [\href{https://arxiv.org/abs/1812.08795}{{\ttfamily
  1812.08795}}].

\bibitem{Cheung:2010gj}
C.~Cheung, G.~Elor, L.~J. Hall and P.~Kumar, \emph{{Origins of Hidden Sector
  Dark Matter I: Cosmology}},
  \href{https://doi.org/10.1007/JHEP03(2011)042}{\emph{JHEP} {\bfseries 03}
  (2011) 042} [\href{https://arxiv.org/abs/1010.0022}{{\ttfamily 1010.0022}}].

\bibitem{Cheung:2010gk}
C.~Cheung, G.~Elor, L.~J. Hall and P.~Kumar, \emph{{Origins of Hidden Sector
  Dark Matter II: Collider Physics}},
  \href{https://doi.org/10.1007/JHEP03(2011)085}{\emph{JHEP} {\bfseries 03}
  (2011) 085} [\href{https://arxiv.org/abs/1010.0024}{{\ttfamily 1010.0024}}].

\bibitem{Chu:2011be}
X.~Chu, T.~Hambye and M.~H.~G. Tytgat, \emph{{The Four Basic Ways of Creating
  Dark Matter Through a Portal}},
  \href{https://doi.org/10.1088/1475-7516/2012/05/034}{\emph{JCAP} {\bfseries
  05} (2012) 034} [\href{https://arxiv.org/abs/1112.0493}{{\ttfamily
  1112.0493}}].

\bibitem{DEramo:2017gpl}
F.~D'Eramo, N.~Fernandez and S.~Profumo, \emph{{When the Universe Expands Too
  Fast: Relentless Dark Matter}},
  \href{https://doi.org/10.1088/1475-7516/2017/05/012}{\emph{JCAP} {\bfseries
  05} (2017) 012} [\href{https://arxiv.org/abs/1703.04793}{{\ttfamily
  1703.04793}}].

\bibitem{Planck:2018vyg}
{\scshape Planck} collaboration, \emph{{Planck 2018 results. VI. Cosmological
  parameters}},
  \href{https://doi.org/10.1051/0004-6361/201833910}{\emph{Astron. Astrophys.}
  {\bfseries 641} (2020) A6}
  [\href{https://arxiv.org/abs/1807.06209}{{\ttfamily 1807.06209}}].

\bibitem{Hryczuk:2021qtz}
A.~Hryczuk and M.~Laletin, \emph{{Dark matter freeze-in from semi-production}},
  \href{https://doi.org/10.1007/JHEP06(2021)026}{\emph{JHEP} {\bfseries 06}
  (2021) 026} [\href{https://arxiv.org/abs/2104.05684}{{\ttfamily
  2104.05684}}].

\bibitem{Tapadar:2021kgw}
A.~Tapadar, S.~Ganguly and S.~Roy, \emph{{Non-adiabatic evolution of dark
  sector in the presence of $U(1)_{L_\mu - L_\tau}$ gauge symmetry}},
  \href{https://arxiv.org/abs/2109.13609}{{\ttfamily 2109.13609}}.

\bibitem{Konar:2021oye}
P.~Konar, R.~Roshan and S.~Show, \emph{{Freeze-in Dark Matter Through Forbidden
  Channel in $U(1)_{B-L}$}},
  \href{https://arxiv.org/abs/2110.14411}{{\ttfamily 2110.14411}}.

\bibitem{2019}
S.~Bhattacharya, P.~Ghosh and N.~Sahu, \emph{Multipartite dark matter with
  scalars, fermions and signatures at lhc},
  \href{https://doi.org/10.1007/jhep02(2019)059}{\emph{Journal of High Energy
  Physics} {\bfseries 2019} (2019) }.

\bibitem{ATLAS:2017oal}
{\scshape ATLAS} collaboration, \emph{{Search for long-lived charginos based on
  a disappearing-track signature in pp collisions at $ \sqrt{s}=13 $ TeV with
  the ATLAS detector}},
  \href{https://doi.org/10.1007/JHEP06(2018)022}{\emph{JHEP} {\bfseries 06}
  (2018) 022} [\href{https://arxiv.org/abs/1712.02118}{{\ttfamily
  1712.02118}}].

\bibitem{CMS:2018rea}
{\scshape CMS} collaboration, \emph{{Search for disappearing tracks as a
  signature of new long-lived particles in proton-proton collisions at
  $\sqrt{s} =$ 13 TeV}},
  \href{https://doi.org/10.1007/JHEP08(2018)016}{\emph{JHEP} {\bfseries 08}
  (2018) 016} [\href{https://arxiv.org/abs/1804.07321}{{\ttfamily
  1804.07321}}].

\bibitem{ATLAS:2017tny}
{\scshape ATLAS} collaboration, \emph{{Search for long-lived, massive particles
  in events with displaced vertices and missing transverse momentum in
  $\sqrt{s}$ = 13 TeV $pp$ collisions with the ATLAS detector}},
  \href{https://doi.org/10.1103/PhysRevD.97.052012}{\emph{Phys. Rev. D}
  {\bfseries 97} (2018) 052012}
  [\href{https://arxiv.org/abs/1710.04901}{{\ttfamily 1710.04901}}].

\bibitem{Curtin:2018mvb}
D.~Curtin et~al., \emph{{Long-Lived Particles at the Energy Frontier: The
  MATHUSLA Physics Case}},
  \href{https://doi.org/10.1088/1361-6633/ab28d6}{\emph{Rept. Prog. Phys.}
  {\bfseries 82} (2019) 116201}
  [\href{https://arxiv.org/abs/1806.07396}{{\ttfamily 1806.07396}}].

\bibitem{MATHUSLA:2018bqv}
{\scshape MATHUSLA} collaboration, \emph{{A Letter of Intent for MATHUSLA: A
  Dedicated Displaced Vertex Detector above ATLAS or CMS.}},
  \href{https://arxiv.org/abs/1811.00927}{{\ttfamily 1811.00927}}.

\bibitem{Tulin:2017ara}
S.~Tulin and H.-B. Yu, \emph{{Dark Matter Self-interactions and Small Scale
  Structure}}, \href{https://doi.org/10.1016/j.physrep.2017.11.004}{\emph{Phys.
  Rept.} {\bfseries 730} (2018) 1}
  [\href{https://arxiv.org/abs/1705.02358}{{\ttfamily 1705.02358}}].

\bibitem{Buckley:2009in}
M.~R. Buckley and P.~J. Fox, \emph{{Dark Matter Self-Interactions and Light
  Force Carriers}},
  \href{https://doi.org/10.1103/PhysRevD.81.083522}{\emph{Phys. Rev. D}
  {\bfseries 81} (2010) 083522}
  [\href{https://arxiv.org/abs/0911.3898}{{\ttfamily 0911.3898}}].

\bibitem{Feng:2009hw}
J.~L. Feng, M.~Kaplinghat and H.-B. Yu, \emph{{Halo Shape and Relic Density
  Exclusions of Sommerfeld-Enhanced Dark Matter Explanations of Cosmic Ray
  Excesses}}, \href{https://doi.org/10.1103/PhysRevLett.104.151301}{\emph{Phys.
  Rev. Lett.} {\bfseries 104} (2010) 151301}
  [\href{https://arxiv.org/abs/0911.0422}{{\ttfamily 0911.0422}}].

\bibitem{Mohapatra:2001sx}
R.~N. Mohapatra, S.~Nussinov and V.~L. Teplitz, \emph{{Mirror matter as
  selfinteracting dark matter}},
  \href{https://doi.org/10.1103/PhysRevD.66.063002}{\emph{Phys. Rev. D}
  {\bfseries 66} (2002) 063002}
  [\href{https://arxiv.org/abs/hep-ph/0111381}{{\ttfamily hep-ph/0111381}}].

\bibitem{1965itpg.book.....V}
W.~G. {Vincenti} and C.~H. {Kruger}, \emph{{Introduction to physical gas
  dynamics}}. 1965.

\bibitem{Colquhoun:2020adl}
B.~Colquhoun, S.~Heeba, F.~Kahlhoefer, L.~Sagunski and S.~Tulin,
  \emph{{Semiclassical regime for dark matter self-interactions}},
  \href{https://doi.org/10.1103/PhysRevD.103.035006}{\emph{Phys. Rev. D}
  {\bfseries 103} (2021) 035006}
  [\href{https://arxiv.org/abs/2011.04679}{{\ttfamily 2011.04679}}].

\bibitem{Kaplinghat:2015aga}
M.~Kaplinghat, S.~Tulin and H.-B. Yu, \emph{{Dark Matter Halos as Particle
  Colliders: Unified Solution to Small-Scale Structure Puzzles from Dwarfs to
  Clusters}}, \href{https://doi.org/10.1103/PhysRevLett.116.041302}{\emph{Phys.
  Rev. Lett.} {\bfseries 116} (2016) 041302}
  [\href{https://arxiv.org/abs/1508.03339}{{\ttfamily 1508.03339}}].

\bibitem{Borah:2021rbx}
D.~Borah, M.~Dutta, S.~Mahapatra and N.~Sahu, \emph{{Singlet-Doublet
  Self-interacting Dark Matter and Radiative Neutrino Mass}},
  \href{https://arxiv.org/abs/2112.06847}{{\ttfamily 2112.06847}}.

\end{thebibliography}\endgroup
\end{document}